\documentclass[%
 reprint,
 amsmath,amssymb,
 aps,
]{revtex4-2}

\usepackage{graphicx}
\usepackage{dcolumn}
\usepackage{bm}
\usepackage{subfigure}
\usepackage{csquotes}
\usepackage{multirow}
\usepackage{subcaption}
\usepackage{float}  
\usepackage{booktabs}
\usepackage{csquotes}

\begin{document}

\preprint{APS/123-QED}

\title{Analysis of quantum-related courses and textbooks for potential integration of quantum sensing}

\author{Namitha Pradeep}
 \email{np6895@rit.edu}
\author{Ben Zwickl}%
\altaffiliation[Also at ]{Center for Computing in Science Education, University of Oslo, Olso Norway}
 
\affiliation{%
 Rochester Institute of Technology, School of Physics and Astronomy, Rochester, NY, USA
}%


\begin{abstract}
The second quantum revolution is driving advancements in quantum computing, communication, and sensing. While quantum computing has gained significant attention in education, quantum sensing remains largely overlooked. In order to find ways of integrating sensing into existing quantum-related curricula, we performed an analysis of six of the most commonly used textbooks in modern physics, quantum mechanics, and quantum computing. We identified a set of keywords related to quantum sensing and tagged all excerpts in which these concepts appeared. For each excerpt, we also recorded the context in which it was presented within the textbook. Network maps were constructed to visualize which keywords appeared in which contexts and the frequency of these occurrences within each subject area. We then developed an analytic rubric to evaluate the conceptual and mathematical depth of these excerpts as well as the extent of sensing-related discussions. Our results show that there is significant variation in how different subjects address these concepts, both in the nature of the content covered and the depth of coverage. We also observe notable differences in how spin-first and position-first textbooks discuss these keywords, particularly in the coverage and contexts in which core concepts such as superposition and entanglement appear. Additionally, we analyzed course titles and descriptions from a database of over 8,000 quantum-related courses, focusing on those that mentioned ``sensing'' or ``sensor''. Together, these analyses of textbook content and course descriptions inform the quantum information science and engineering education community about potential opportunities to integrate quantum sensing topics into quantum-related courses in physics and adjacent disciplines. 
\end{abstract}

\maketitle

\section{Introduction}

The ability to effectively manipulate quantum systems and to harness unique quantum properties such as superposition and entanglement in the development of quantum technologies has led to what is often referred to as the ‘second quantum revolution’ \cite{macfarlane_quantum_2003}.
As the quantum industry continues to grow, there is an increasing need for talent from all levels to lead the innovation, design, manufacturing, and application of these technologies. \cite{masiowski_quantum_2022,nstc_quantum_2022}.
While projections indicate an exponential increase in quantum technology jobs over the next two decades, companies have reported increasing difficulty in finding candidates with the necessary skill sets \cite{venegasgomez_quantum_2020}.
In response to this emerging quantum revolution, there has been a call to develop educational initiatives aimed at expanding the quantum workforce in many countries around the world, for example, the Quantum Flagship project in the EU \cite{noauthor_homepage_nodate} and the National Quantum Initiative in the US \cite{rep_smith_hr6227_2018}.

The technologies of the second quantum revolution are often categorized into three `pillars’: quantum computing, quantum networking and communication, and quantum sensing, which is our focus.
Quantum sensing refers to the use of quantum systems such as atoms, photons, or superconducting circuits, and quantum phenomena such as superposition or entanglement to measure a physical quantity
Using these uniquely quantum properties like entanglement, quantum sensors can achieve measurement sensitivities that surpass classical sensors.
This is a rapidly growing field, with its real-world applications such as quantum gravimeters or NV centers for bio-sensing are becoming accessible much sooner than practical applications of quantum computing \cite{degen_quantum_2017}.
Acknowledging the near-term economic and technological impact of quantum sensors, efforts at the national level encourage agencies to prioritize and develop new quantum sensing approaches \cite{nstc_bringing_2022}.
The European Competence Framework for Quantum Technologies \cite{directorate-general_for_communications_networks_content_and_technology________________________________________european_commission_european_2025} has identified Quantum sensors and Imaging systems as one of the four quantum technology application domains.
It is also one of the eight core domains in the framework, which is a taxonomy of potential knowledge and skills required in the quantum technology workforce.
Despite this, quantum sensing has received the least attention in the development of curriculum resources and in quantum technology courses.
Educators have observed `curricular breadth gaps' at the course level, wherein quantum sensing gets very limited emphasis \cite{el-adawy_insights_2025} and most courses tend to focus more on quantum computing \cite{meyer_todays_2022}.
In a study of 63 quantum information science and engineering (QISE) courses at the postsecondary level, it was found that only 22\% addressed quantum sensing-related topics, and none were dedicated courses on quantum sensing \cite{meyer_introductory_2024}.
Experts in academia and industry have reported feeling less competent in quantum sensing compared to other areas of QISE \cite{greinert_future_2023}.
Most quantum mechanics and quantum computing textbooks currently used in the physics curriculum emphasize quantum theory \cite{griffiths_introduction_2018,mcintyre_quantum_2022,mermin_quantum_2007,bernhardt_quantum_2019}, and there are no textbooks focusing on hardware implementations and applications of quantum technologies such as quantum sensing at the undergraduate level\cite{asfaw_building_2022}.
Educators have also reported difficulty accessing primary literature on these topics, as they are often non-experts in the field \cite{meyer_todays_2022}.

Quantum sensing is challenging from an educational perspective as there are no established conceptual frameworks or standard textbooks, and its applications span multiple fields.
Different sensing platforms use very different hardware and operating principles, and hence understanding these systems requires a diverse set of concepts.
Due to this diversity, students need to build interdisciplinary background knowledge, including ideas from physics, engineering, chemistry and nanoscience.


Given this relative lack of emphasis on quantum sensing in current curricula, there is an opportunity to better integrate this area into quantum education.
The long-term goal of our project is to develop modular curricular resources that can be incorporated into existing undergraduate level quantum-related courses.
An initial step toward this goal is to understand the inclusion of quantum sensing and associated concepts within the wider landscape of current quantum curricula in higher education.
By characterizing the topics currently being addressed in undergraduate quantum-related textbooks and the concepts covered in quantum-related courses, we can better understand the extent of students' existing knowledge and identify natural entry points in the curriculum where sensing-related topics can be introduced.
This, in turn, will enable us to build on the existing curriculum and develop suitable learning objectives and learning progressions for these courses.

In addition to this, examining existing courses that address sensing-related topics provides insight into the current curricular landscape.
Identifying the disciplines these courses are offered under would help to understand where quantum sensing modules could be meaningfully integrated.
It is also important to know the types of content covered in these courses as this informs the kind of resources that would be appropriate for integration.
Different course types could have distinct content coverage, emphases (e.g., theoretical, experimental, or computational) and learning outcomes, and therefore require differently tailored curricular resources.

To this end, we conducted a (1) a textbook analysis to examine how sensing-related concepts are addressed in commonly used textbooks, and (2) a course analysis to investigate  the types of courses that include quantum sensing topics. We are trying to answer the following research questions:
\begin{itemize}
    \item Across 6 textbooks in quantum mechanics, modern physics, and quantum computing that are widely used in undergraduate physics programs:
    \begin{enumerate}
        \item How do these textbooks and subjects differ in the extent and depth in which they address keywords related to quantum sensing?
        \item In which broad physics topics (contexts) do these quantum sensing-related keywords appear in these textbooks?
        \item How do the sensing keywords and the physics topics addressed  differ between textbooks that start with two-state systems vs position basis wavefunctions?
    \end{enumerate}
    \item Across STEM, 
    \begin{enumerate}
    \setcounter{enumi}{3}
        \item What kinds of courses are already addressing quantum sensing?
        \item How are courses mentioning ``quantum'' and ``sensing''/``sensor(s)'' distributed across disciplines and course types?
    \end{enumerate}
\end{itemize}

\section{Background}

\subsection{Overview of research on quantum and QISE curricula}

A recent study focused on quantum education across all disciplines, not just physics, identified more than 8000 quantum-related courses in the United States.
Over 4700 courses were identified in physics departments that mentioned the word ``quantum'' in the course title or description.
Of these, approximately 1600 were quantum mechanics (QM) courses (QM 1, QM 2, and graduate-level QM courses), and 600 were modern physics (MP) courses \cite{pina_landscape_2025}.
Since QM and MP appear to be the courses where students most often encounter quantum concepts, we selected these subjects for our textbook analysis.

In the same study, over 500 courses were identified as dedicated QISE courses across all disciplines, more than 200 of which were taught in physics departments.
However, the vast majority of QISE courses (418 of 514) focused on quantum computing and information, with very few courses addressing other aspects of QISE.
This trend is reflected in other studies mentioned in the introduction \cite{meyer_todays_2022,meyer_introductory_2024}.
Thus, we justify quantum computing (QC) as a 3rd common type of quantum course for inclusion in our textbook analysis.
Also, this quantum education landscape study \cite{pina_landscape_2025} provided the database of quantum-related courses needed to address research questions 4 \& 5. 

Buzzell et al.’s study of QM course syllabi from 188 institutions showed that all institutions required physics majors to take at least one quantum course.
The most frequently taught topics in these courses were the Schrödinger equation and wavefunctions in three dimensions.
In contrast, the Stern–Gerlach experiment appeared in only about 26\% of course syllabi, even though it provides a way to introduce a wide range of concepts using the formalism of two-state systems, which also connects to QISE topics (e.g., qubits)\cite{buzzell_quantum_2025}.
We use this syllabi analysis to identify the most commonly used quantum textbooks.

In a separate study of 167 MP course syllabi, 94\% of the courses were found to cover quantum topics, including the photoelectric effect, wave–particle duality, and the Schrödinger equation.
Physics majors are the primary audience for these courses, which they typically take in their second year, with Newtonian physics , Electricity \& Magnetism, and Calculus II as prerequisites at most of the institutions \cite{buzzell_modern_2025}.
MP courses are often  students’ first introduction to quantum concepts.
We use this syllabi analysis to identify the most commonly used MP textbooks.

\subsection{Textbook analysis}

Valverde et al. \cite{valverde_according_2002} proposed a modified tripartite model of curriculum in which textbooks and other resource materials serve as a link between the intended curriculum and the implemented curriculum, referred to as the `potentially implemented curriculum'.
Given this central role textbooks play in shaping what students learn, analyzing them can inform the development of evolving curricula \cite{okeeffe_framework_2013}.
Textbook analysis has been used to examine content coverage, topic placement, and the development of key concepts across disciplines such as mathematics \cite{son_what_2017} and engineering \cite{purzer_teaching_2010}.
In physics, studies have explored how textbooks address specific topics, such as weight \cite{taibu_textbook_2015} and the nature of the electromagnetic field \cite{suarez_electromagnetic_2023}.
In this study, we conduct a content analysis of textbooks to identify meaningful entry points for integrating quantum sensing curricular resources.
This analysis also helps us determine students’ likely prior knowledge, enabling us to develop coherent learning progressions.

\subsubsection{Spin-first vs position-first approaches}

There are two common instructional approaches used in quantum mechanics courses - the spin-first approach, which begins with two-level systems, and the position-first approach, which starts with wave mechanics and the Schrödinger equation.
The position-first approach (also called wavefunction-first) is the more traditional approach used in QM classes, and uses the wavefunction formalism in the position basis, and typically covers topics such as particle in a box or potential wells.
The spin-first approach uses the Dirac notation and covers topics such as physics of spin-1/2 systems and Stern-Gerlach experiment.
In Buzzell et. al’s study, it was found that all MP courses used the position-first approach and only 26\% of the QM courses used the spin-first approach.
The choice of textbook correlated with instructors' decisions to use a position-first or spin-first approach. All courses that used David Griffiths' \textit{Introduction to Quantum Mechanics} \cite{griffiths_introduction_2018} adopted a position-first approach, while all courses using David McIntyre's \textit{Quantum Mechanics: A Paradigms Approach} \cite{mcintyre_quantum_2022} adopted a spin-first approach \cite{buzzell_quantum_2025}.
Since the choice of textbook is linked to the instructional approach and can potentially influence the topics covered, a textbook analysis and comparison across textbooks becomes highly relevant, particularly through the lens of spin-first vs position-first approaches, which is addressed in research question 3.

\section{Methods}

\subsection{Textbook analysis}
In this study, we identified three key subjects - quantum mechanics, quantum computing, and modern physics - as the most appropriate for the inclusion of quantum sensing modules.

\subsubsection{Textbook selection}
We selected two of the most widely used textbooks for undergraduate courses in each subject for our analysis.
For quantum mechanics, we referenced \citet{buzzell_quantum_2025}, according to which \textit{Introduction to Quantum Mechanics} by D. J. Griffiths \cite{griffiths_introduction_2018} is the most commonly used textbook, adopted by 56\% of institutions, which was therefore selected for analysis.
This was followed by \textit{Quantum Mechanics: A Paradigms Approach} by R. McIntyre \cite{mcintyre_quantum_2022} and \textit{A Modern Approach to Quantum Mechanics} by J. S. Townsend \cite{townsend_modern_2012}, each used in 13\% of institutions. 
Both textbooks are viable choices, but we selected McIntyre due to its greater recognition within the PER community and our familiarity with its content.

For modern physics (MP), we used the data on textbook usage that the authors of \cite{buzzell_modern_2025}, a study on 167 modern physics courses in the U.S., shared with us.
The most commonly used textbooks were found to be \textit{Modern Physics} by Kenneth S.Krane \cite{krane_modern_2012} (adopted in 22 courses) and \textit{Modern Physics} by  R. A. Serway, C. J. Moses, and C. A. Moyer \cite{serway_modern_2005} (adopted in 14 courses), both of which were selected for analysis. There was more diversity in textbook choice for MP courses compared to QM.

For quantum computing, Meyer et al. (2022) \cite{meyer_todays_2022}, a survey of quantum information science instructors, indicated a lack of consensus on a standard textbook.
The study identified 14 different textbooks across 27 courses, with \textit{Quantum Computer Science: An Introduction} by N. D. Mermin \cite{mermin_quantum_2007} used in 7 courses and \textit{Quantum Computation and Quantum Information} by Nielsen and Chuang \cite{nielsen_quantum_nodate} used in 6 courses.
We selected Mermin’s textbook alone, as Nielsen and Chuang’s text is more advanced and better suited for graduate-level discussions.
Additionally, \textit{Quantum Computing for Everyone} by Chris Bernhardt \cite{bernhardt_quantum_2019} was included, as it is also used and recommended as a beginner-friendly quantum computing textbook \cite{galetto_experience_2024,german_quantum_2023,suits_electronics_2023}.
For consistency, all textbooks will be referred to by the last name of the first author throughout this paper.

\subsubsection{Keyword selection}
To understand how these textbooks address sensing-related content, we identified a set of keywords representing general concepts related to sensing and analyzed how these keywords are included in each textbook.
This method of using keywords ensures a structured approach to examining multiple textbooks.
Also, the keywords help pinpoint sections where sensing-related topics are discussed, allowing for a focused analysis without the need to manually review entire textbooks.

\citet{degen_quantum_2017} is a review article that introduces the fundamental principles, methods, and applications of quantum sensing.
This article presents various experimental implementations of quantum sensors, most of which were incorporated into our keyword list (e.g., neutral atoms, Rydberg atom, trapped ion, NMR, SQUID, optomechanics, and interferometer).
Additionally, we conducted a word frequency analysis of the article to identify the most commonly occurring concepts.
From the list of most frequent words, the first 20 were directly included in our initial selection of keywords (signal, sensing, sensor, noise, sensitivity, sequence, estimation, readout, detection, pulse, (de)coherence, optical, squeezing, entanglement, (de)coupling, control, interference, clock, spectroscopy, and Ramsey).
From the remaining words in the list, we selected additional relevant terms, including specific hardware platforms (e.g., superconduct-), applications (e.g., magnetometry, imaging), techniques (e.g., relaxometry, resonance), and a few generic terms (e.g., metrology, probe, and device).

Another source for identifying keywords was the Competence Framework for Quantum Technologies \cite{directorate-general_for_communications_networks_content_and_technology________________________________________european_commission_european_2025}.
Several sub-items in the Quantum Sensing domain in the framework correspond to relevant sensing applications, which were added to the keywords list (e.g., electromagnetic field sensor, temperature sensor, gravitational wave detector or gravimeter, quantum imaging, and atomic clock).
Apart from these sources, we selected a few keywords related to general concepts that are important in this context (e.g., measurement, photonics, and laser cooling).

We made some modifications to certain keywords.
For example, interferometry and interferometer are both important concepts that occur in similar discussions, so we combined them into one keyword: interferometer/interferometry.
Sections mentioning either term were recorded under the same keyword.
Similar adjustments were made for keywords such as detector/detection, estimation/estimator, and nanoscale/nanosensing/nanotech.
For some keywords, we wanted to capture all variations of a word, so we used keywords like superconduct---, which includes instances of superconductors, superconducting, superconductivity, superconductive, etc.
Similarly, we used the keyword entangle--- to also cover entangled and entanglement.

The first round of keywords included 74 words, but after conducting a preliminary analysis by reviewing the keywords in the textbooks, we found that many keywords did not appear anywhere in the books, so these were removed from the list and not considered for further analysis (for example, Ramsey, shot noise, readout, etc.).
Among the remaining ones, some were addressed only in problems, footnotes, or appendices, rather than in the main text (e.g., tomography).
Some others appeared predominantly in very generic and not so relevant excerpts since these could refer to multiple concepts (e.g., control, sequence, device, etc.).
So these keywords were also removed from the list.
However, we retained a small subset of these, whose absence from the textbooks was important to highlight in itself (e.g., NV center, gravimeter, photonics, etc). 
After these revisions, a final list of 41 keywords was established.
These keywords can be grouped based on the type of sensing-related concept they represent (see Table  \ref{tab:keywords}).
For the sake of brevity and interest, we discuss only the Core Concepts, Techniques/Applications, and Hardware categories in this paper.
A detailed discussion of results for the General category of keywords can be found in the Supplemental Material \cite{supp}.

\begin{table}[h]
    \centering
    \begin{tabular}{|p{3cm}|c|}
        \hline
        Category & Keywords \\ 
        \hline
        \multirow{4}{3cm}{Core concepts} & Superposition \\
                                                               & Interference \\
                                                               & Entangle--- \\
                                                               & Measure--- \\
        \hline
        \multirow{5}{3cm}{Techniques / Applications} & Imaging \\
                                      & Interferometer \textnormal{or} Interferometry \\
                                      & Laser cooling \\
                                      & Magnetometry \textnormal{or} Magnetometer \\
                                      & Relaxometry \\
                                      & Resonance \\
                                      & Spectroscopy \\
                                      & Atomic clock \\
                                      & Chemical sensing \textnormal{or} sensor \\
                                      & Electromagnetic field sensor \\
                                      & Gravimeter \\
                                      & Gravitational \textnormal{or} Gravity wave detector\\
                                      & Gyroscope \\
                                      & Navigation \\
                                      & RADAR / LIDAR \\
                                      & Temperature sensor \textnormal{or} Thermometer \\
        \hline
        \multirow{17}{3cm}{Hardware} & Superconduct--- \\ 
                                       & Josephson junction \\
                                       & SQUID \\
                                       & NMR \textnormal{or} nuclear magnetic resonance \\
                                       & NV center \\
                                       & Nanoscale / Nanosensing / Nanotech \\
                                       & MEMS - micro electro-mechanical sensor \\
                                       & Optomechanic \\
                                       & Photonics \\
                                       & SET - single electron transistor \\
                                       & Quantum dot \\
                                       & Trapped ion \\
        \hline                               
        \multirow{10}{3cm}{General} & Quantum technology \\
                                      & Quantum imaging \\
                                      & Quantum sensor \\ 
                                      & Sensor \\ 
                                      & Sensing \\ 
                                      & Metrology \\
                                      & Detector / Detection \\
                                      & Estimation / Estimator \\
                                      & Probe \\
        \hline
        
    \end{tabular}
    \caption{List of all 41 keywords grouped according to the category of keyword}
    \label{tab:keywords}
\end{table}

\subsubsection{Data collection}
The data collection process involved searching for these keywords in each textbook and recording every excerpt where a keyword appeared.
An excerpt is a portion of text where a keyword appears and includes a single cohesive discussion of the topic/theme being addressed in that portion, where the topic/theme of the excerpt we consider is independent of the keyword itself.
Additional sentences (from before and/or after the selected portion) may also be considered part of the excerpt for forming a unit with sufficient meaning, as long as they fit into the theme addressed in that segment.
In cases where the keyword is distributed across multiple pages within a single unified discussion, the entire segment is considered a single excerpt.
In addition to the excerpts from the main body of text, we also recorded problems, footnotes, and worked examples as individual excerpts if they contained the keyword.
Also, the same excerpt could be associated with multiple keywords, in which case it was recorded separately under each keyword.

For every excerpt, we documented its chapter number, chapter name, and section name.
We also recorded the “context,” which refers to the broader conceptual category under which the discussion falls in that textbook.
A list of contexts was identified (common across all textbooks) and continuously revised as more excerpts were recorded, following an inductive approach.
For example, an excerpt tagged with the keyword Imaging might discuss the scanning tunneling microscope, which falls under the context of Quantum Tunneling in the textbook.
So Quantum Tunneling was added to the `list of contexts' if it was not already included.
Our final list consisted of 66 different contexts, which were used for network analysis, as discussed in section \ref{subsec:data_analysis_network}.

\subsubsection{Excerpt-level analysis: Rubric and Heatmaps}\label{subsec:Method_rubric_heatmap}
A rubric was developed to evaluate each excerpt based on how it addresses the keywords and discusses sensing or other relevant concepts.
The rubric consists of two broad categories with multiple criteria within each, and we assigned each excerpt a score on a predefined scale for each criterion.
This approach ensures consistency in the analysis and facilitates a quantitative comparison across the textbooks.

\paragraph{\textbf{Structure of the rubric -}}

\begin{figure}[hbtp]
\includegraphics[scale=0.38]{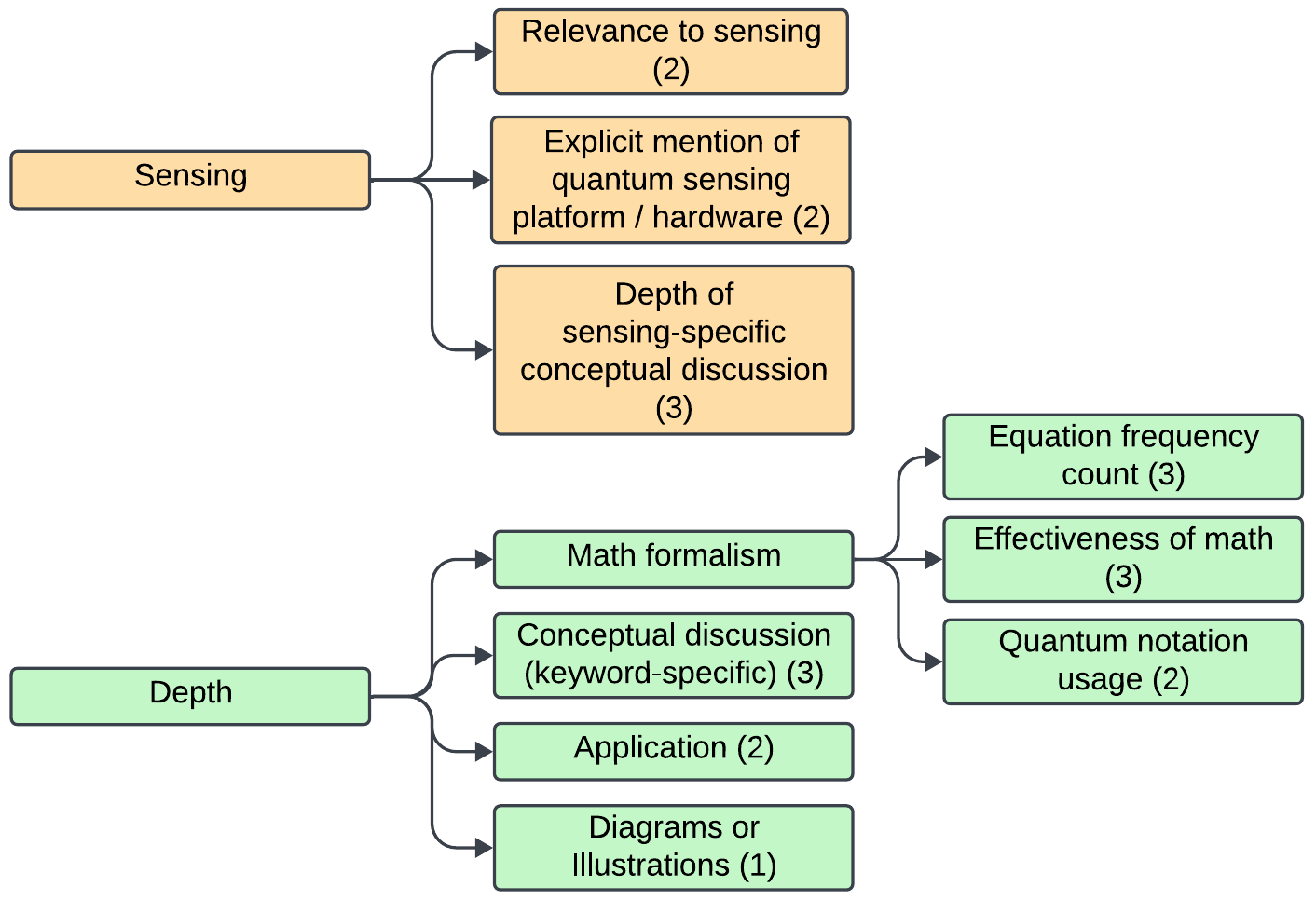}
\caption{\label{fig:rubric}Structure of the scoring rubric. The rubric consists of two categories - Sensing and Depth - with multiple criteria within each. The number in the parenthesis denotes the maximum score for that criterion}
\end{figure}

Our goal was to analyze two main aspects of how the keywords were included within the textbooks: the sensing-related content and the depth of the keyword-related discussion.
These two aspects became the main categories in our rubric (Fig. \ref{fig:rubric}).

\subparagraph{Sensing} Within the sensing category, we have three sensing criteria:
\begin{itemize}
  \item \textit{Relevance} to sensing (scored on a scale of 0–2)
  \item \textit{Sensor platform or hardware} is explicitly mentioned (scored on a scale of 0–2)
  \item Depth of \textit{sensing-specific conceptual discussion} (scored on a scale of 0–3)
\end{itemize}
The \textit{Relevance} criterion evaluates whether the excerpt is directly related to sensing or addresses a concept that is adjacent to sensing. This criterion is scored on a scale from 0 to 2.
For example, the excerpt could explicitly discuss sensing (level 2), be related to understanding basic principles of sensing (level 1), or be not directly relevant to sensing-related concepts (level 0).

The \textit{Sensor platform or hardware} criterion assesses whether the excerpt addresses a quantum sensing platform (e.g., atom interferometer, atomic clock, etc.), mentions a classical sensing platform, or does not mention any hardware at all.
This helps identify whether the excerpt discusses actual devices and implementations or remains at an abstract level.
It is also scored on a scale from 0 to 2.

The \textit{sensing-specific conceptual discussion} criterion evaluates the level of detail and complexity in the conceptual explanation of a sensing protocol (a sequence of steps used to detect a physical quantity using a given sensing platform).
The scoring criteria range from no mention of sensing concepts to a comprehensive, detailed description of how the sensor works and its experimental implementation, and is scored on a scale from 0 to 3.

\subparagraph{Depth} In the depth category, we included six criteria that evaluate multiple aspects of how comprehensive the keyword-focused discussion is, including mathematical formalism, conceptual depth, use of diagrams, and mentions of applications.
The six depth criteria are:
\begin{itemize}
  \item \textit{Equation frequency count} (scored on a scale of 0-3)
  \item \textit{Effectiveness} of mathematical representation in conveying the concepts (scored on a scale of 0-3)
  \item \textit{Quantum notation usage} (scored on a scale of 0-2)
  \item Depth of \textit{keyword-specific conceptual discussion} (scored on a scale of 0–3)
  \item \textit{Application} (scored on a scale of 0-2)
  \item \textit{Diagrams} or illustrations (scored on a scale of 0-1)
\end{itemize}

The \textit{Equation frequency count} assesses the number of mathematical equations in the excerpt, providing insight into the mathematical rigor of the discussion.
The scoring levels range from no equations (0), to 1-2 equations (1), to more than two equations with a reasonable mathematical description (2), and finally a complete derivation or explanation of a mathematical model (3), making it a scale of 0-3.
When determining the equation frequency count score, only equations specifically relevant to the keyword are considered.

The \textit{Effectiveness} criterion evaluates how well the mathematical formalism in the excerpt captures the concepts (keyword-specific content) being discussed.
It is scored on a scale of 0-3, where a low score indicates that the excerpt covers many different ideas but without much mathematical detail.
A high score indicates that the equations actively contribute to capturing all the keyword-related ideas in the excerpt, providing a complete mathematical explanation of the concept.

The \textit{Quantum notation usage} criterion assesses the extent to which quantum-specific mathematical formalism, such as Dirac notation, qubit notation, and wavefunction representations, is used.
It is scored on a scale of 0-2, where 0 indicates no quantum notation, 1 indicates quantum notation mentioned in the text but not in equations, and 2 indicates at least one equation using quantum formalism.
Again, the mathematical formalism is considered for scoring only if it is directly relevant to the keyword-specific content.

The \textit{keyword-specific conceptual discussion} criterion measures how thoroughly the excerpt explains the concept related to the keyword, and is scored on a scale of 0-3.
A low score (0) indicates that the excerpt does not engage with the concept meaningfully, while a moderate score (1-2) suggests a brief discussion or partial explanation.
A high score (3) signifies that the excerpt provides an in-depth conceptual explanation of the concept, potentially including fundamental principles, detailed explanations of the underlying mechanisms, and connections to other topics.

The \textit{Application} criterion evaluates if the excerpt mentions applications of the keyword-focused concept, which includes scientific and technological applications, real-world applications (for example, medical imaging or navigation), and detailed descriptions of how the application works, and the score ranges from 0-2.

Finally, the \textit{Diagrams} criterion evaluates whether the textbook uses diagrams, illustrations, or figures, to visually explain the keyword-related concepts in the given excerpt.
This criterion is scored on a binary scale of 0-1.

\paragraph{\textbf{Applying the rubric -}}

The next step involved applying the rubric to the excerpts and assigning a score for each criterion.
Once the excerpt and all relevant details have been identified, we assigned a score to the excerpt in each of the sensing and depth criteria based on the rubric.
It is important to note that when analyzing the excerpt for the sensing criteria, we consider the general sensing content within the excerpt.
However, for the depth criteria, we specifically assess the mathematical formalism, conceptual depth, diagrams, and applications solely in relation to the keyword, rather than the broader content of the excerpt.
This ensures that the depth scores reflect the depth of keyword-specific content rather than the broader contextual content of the excerpt.

Problems, footnotes, and end-of-chapter summary points generally received a score of zero for all criteria.
There were rare exceptions when a back-of-the-chapter problem is highly detailed, provides the necessary conceptual background to approach the problem, or thoroughly discusses an application.
Although excerpts from problems, footnotes, and summaries did not significantly impact the overall keyword scores (next section), the contexts in which they appeared were recorded and incorporated into the network analysis (see Section \ref{Res:network}).

\paragraph{\textbf{Sensing scores, Depth scores, and Heatmaps -}}
Using the scores assigned for individual criteria, we calculated an excerpt sensing score and an excerpt depth score for each excerpt.
These were then used to determine a final keyword sensing score and a final keyword depth score for every keyword in each textbook.

\subparagraph{Sensing score calculation} After applying the rubric to all excerpts, the excerpt sensing score (the sensing score for each individual excerpt) is determined by summing the scores from the three sensing criteria: \textit{Relevance to sensing} (scored on a scale of 0–2), mention of \textit{sensor platform or hardware} (scored on a scale of 0–2), and depth of \textit{sensing-specific conceptual discussion} (scored on a scale of 0–3).
Thus, the sensing score for the $j$-th excerpt is calculated as the sum of the scores assigned to the excerpt in each of the three criteria: 
\begin{equation*}
  \textrm{$s_{j,k}$} = \sum\limits_{i ~ \textrm{in sensing criteria}} s_{i,j,k}   
\end{equation*}
where $s_{j,k}$ represents the excerpt sensing score of excerpt $j$ for keyword $k$, and $s_{i,j,k}$ is the sensing sub-score for criterion $i$ and excerpt $j$ for keyword $k$.
Each excerpt can hence receive a maximum sensing score of 7. 
The final sensing score for a keyword $k$ (keyword sensing score $S_k$), as displayed in the heatmap, is obtained by summing all the individual excerpt sensing scores for that keyword:
\begin{equation*}
    \textrm{$S_k$} = \sum\limits_{\textrm{$j$}} \textrm{$s_{j,k}$}
\end{equation*}
So the maximum possible score for each keyword depends on the number of excerpts, and it is given by the number of excerpts times 7.

Summing the sensing scores across excerpts provides a holistic measure of how extensively a textbook addresses sensing-related keywords, and reveals which keyword receives the most attention in a textbook.
A higher keyword sensing score generally suggests more substantial coverage across multiple sections of the book.
At the same time, it is important to remember that a lower keyword score does not always mean the excerpts did a poor job of covering sensing-related content.
For example, the keyword SQUID has a keyword sensing score of 6 in Krane, which is low on the total scale.
However, this score comes from a single excerpt, meaning that excerpt scored 6 out of 7, which is a strong sensing score.
This highlights a limitation of looking only at the total sum.
A keyword with many excerpts, each with low sensing content, could end up with a higher total score than a keyword with just one excerpt that provides a rich discussion.

\subparagraph{Depth score calculation} The excerpt depth score is calculated by summing the individual scores for the depth criteria - \textit{Equation frequency count} (scored on a scale of 0-3), \textit{Effectiveness} (scored on a scale of 0-3), \textit{Quantum notation usage} (scored on a scale of 0-2), \textit{Keyword-specific conceptual discussion} (scored on a scale of 0–3), \textit{Application} (scored on a scale of 0-2), and \textit{Diagrams} (scored on a scale of 0-1).
\begin{equation*}
   d_{j,k} = \sum\limits_{\textrm{\textit{i} in depth criteria}} d_{i,j,k} 
\end{equation*}
where $d_{j,k}$ represents the depth score of excerpt $j$ for keyword $k$, and $d_{i,j,k}$ is the depth sub-score for criterion $i$ and excerpt $j$ for keyword $k$.
The maximum possible excerpt depth score is 14.

The keyword depth score ($D_k$), as shown in the heatmaps, represents the highest depth score achieved by any individual excerpt for that keyword.
Unlike the keyword sensing score, which is calculated as the sum of all excerpt scores, the keyword depth score is based solely on the highest-scoring excerpt, which we refer to here as the ``best excerpt" score.
This approach is used because we are interested in a measure of how thoroughly a concept is explained,  so using the best excerpt score highlights the strongest discussion that the textbook offers for a particular keyword.
Also summing all excerpt depth scores could give a misleadingly high score for a keyword that appears frequently but only in shallow discussions.
\begin{equation*}
   D_k = \max\limits_{j} \, \, d_{j,k} 
\end{equation*}
Consequently, the maximum possible keyword depth score remains 14, identical to the maximum excerpt depth score.

\subparagraph{Heatmaps} The final keyword scores were visualized in heatmaps (Fig. \ref{fig:hm_phenomena}, for example) that display sensing and depth scores using a color mixing scheme. 
Keyword sensing scores are normalized to a 0-1 range and represented using a red color scale, ranging from black (0) to bright red (1).
Similarly, keyword depth scores are normalized to a 0-1 range and represented using a blue color scale, ranging from black (0) to bright blue (1).
Each cell in the heatmap, corresponding to a specific keyword–textbook pair, displays a blended color formed by combining the red intensity (sensing score) and blue intensity (depth score). So, for example, bright magenta (which is a combination of bright red and bright blue) indicates a high sensing and high depth score.

\subparagraph{Comment on excerpts containing multiple keywords.} There are instances where the same excerpt is tagged under multiple keywords when those keywords are all present within that section.
For example, a discussion on Josephson junctions could be associated with the keywords Josephson junction, SQUID, interferometry, ‘superconduct—,’ and maybe resonance.
So it is essential to clarify how the sensing and depth scores differ in these cases.
For the excerpt sensing score, the evaluation considers the overall sensing-related content in the excerpt and whether it explicitly mentions any sensing platform or hardware, regardless of the keyword under which it was tagged.
As a result, cross-listed excerpts receive the same excerpt sensing score across all instances.
However, the depth criteria are specific to the individual keyword.
For example, when assessing equation frequency or conceptual discussion depth, the focus is solely on equations or conceptual explanations relevant to that specific keyword, and not the general content of the excerpt.
For example, an excerpt may contain equations related to the keyword SQUID (if the equations pertain to the device) but have an equation frequency count of zero for the keyword superconduct—.
This condition applies to all depth criteria, and hence, the same excerpt can receive different excerpt depth scores when cross-listed under different keywords.

\subsubsection{Network analysis of excerpts in context} \label{subsec:data_analysis_network}

To analyze the contexts in which different keywords appear and their frequency, we constructed a network representation.
In this network, we have nodes representing both keywords and contexts, while edges indicate occurrences of a keyword within a specific context.
Each time an excerpt for a particular keyword appears in a given context, an edge is drawn between the corresponding keyword node and context node.

To ensure consistency in identifying the context of each excerpt, we checked the intercoder reliability.
A subset of 43 excerpts, selected from all six textbooks, was independently coded for context by two additional coders.
Prior to any discussion, all three coders had perfect agreement on the context for 19 excerpts.
For 6 excerpts, the coders identified similar contexts, differing only in the level of generality (for example, Deutsch’s algorithm vs Quantum algorithms).
In these cases, we resolved discrepancies by agreeing on a common level of generality, corresponding to the section in which the excerpt was located.
For 16 excerpts, two coders agreed while the third proposed a slightly different context (for example, quantum transitions vs laser excitations).
And only for 2 excerpts did all coders identify slightly different contexts.
These 18 excerpts were discussed in detail and we reached a final consensus on all of them.
It was noted that excerpts concerning quantum measurements and the double-slit experiment were particularly challenging, as these topics frequently appeared in sections that addressed multiple concepts or were presented in diverse contexts.
Based on this insight, the primary coder conducted a review of all excerpts involving these topics.
During discussion, we also recognized that some excerpts were best classified under multiple contexts.
For example, an excerpt discussing superposition states in a quantum well in the textbook section ‘Superposition states and Time dependence’ was classified under two contexts: ``Quantum wells" and ``Time evolution".
To address this, we systematically reviewed the section titles of all excerpts and flagged cases where multiple concepts might be present within the section.
These excerpts were re-coded under more than one context, if found appropriate.
It is worth noting that the list of contexts applied to the excerpts was common across all textbooks and was not unique to each. 

\subsection{Course analysis}

In order to address research questions 4 and 5, we analyzed courses that included the terms ``quantum" and ``sens—" in their title or description across multiple disciplines.

\subsubsection{Course data collection}

The course data used in this study was collected as part of another project conducted by our group to examine the educational landscape of quantum information science in the U.S. \cite{pina_landscape_2025}.
A team of eight members systematically searched the program and course catalogs of 1456 institutions nationwide, identifying all courses that included the word ``quantum" in their title or course description.
For each course that met this criteria, the course title, course number, description, prerequisites, associated departments, and course level were recorded.
A total of 8456 courses were recorded and are available in an interactive online dataset \cite{streamlit}.

From this dataset, we filtered out and selected those courses that contained variations of ``sens—" (such as sensing or sensor) in their title or description.
These courses will hence contain both the words ``quantum" and ``sens—" in their descriptions (not necessarily together), and with this, we expect to capture courses that have some focus on quantum sensing topics.
A total of 121 courses from 76 different institutions were identified in this subset.

\subsubsection{Course data analysis}

To examine the content of these courses, we analyzed the course descriptions to qualitatively assess the extent to which the content was related to sensing, focused on QISE, or connected to QISE-adjacent topics.
From this analysis, we identified 12 different thematic groupings (categories) based on the topics they covered.
Within each category, we also identified the types of sensors or sensing applications explicitly addressed in the descriptions, for example, nanobiosensors, remote sensing, accelerometry, etc.
Additionally we looked for occurrences of the keywords used in the textbook analysis within the course descriptions and examined how they were addressed.

An intercoder reliability check was performed to ensure the codebook could be consistently applied for categorizing courses.
Two coders independently coded two rounds of data, with each round consisting of around 20\% of the dataset (25 courses each). 
In the first round, there was 50\% agreement prior to discussion.
Most disagreements were due to the nuanced definitions of the categories.
For example, the distinction between the categories `Quantum Engineering and Technologies' and `Quantum Computing and Information' (more details on the categories are discussed in section \ref{Res:course}) was subtle, as both involved similar content, with the former requiring an explicit mention of hardware or physical implementation.
Hence, a few category definitions (for categories `Quantum Engineering and Technologies', `Quantum Computing and Information', and `QIST adjacent') were updated.
For the categories `Devices - theory and design' and `Solid state and Materials science', a few additional topics were added to the category definition. 
The codebook was thus revised to provide clearer inclusion criteria and clearly distinguish between categories. 
Using the revised codebook, the coders independently conducted another round of coding, and there was 80\% agreement prior to discussion.
Discrepancies arose when course descriptions could potentially be assigned to more than one category.
For example, a couple of courses addressing optical sensing could be classified as either `Optics/Photonics' or `Metrology/Sensing' because the descriptions were ambiguous and included content related to both areas.
These cases were discussed by the coders, who agreed to assign each course to the topic area that the course seems to lean more heavily on.
Consensus was reached after discussion and the codebook was not revised any further.

The final categorization and a detailed analysis of the types of content covered in each category will be presented in the Results section (section \ref{Res:course}).
We also analyzed the distribution of these courses across the 12 categories and the departments in which they were offered.

\section{Results: Textbook Analysis}\label{TB_overview}
\subsection{Examples of high and low scoring excerpts}\label{Res:Rubric}

We first examine some examples of excerpts that received high and low excerpt sensing scores. One example of a high-scoring excerpt is from McIntyre (Chapter 6 - Unbound States, pg. 196), which was tagged under multiple keywords: interference, interferometer, gyroscope, and navigation.
\begin{quotation}
    ``One of the important features of an atom interferometer is its ability to measure extremely small changes in potential energy. This ability arises from the dependence of the de Broglie wavelength of the particle on the potential energy ... \\ ... A measurement of the potential energy with an atom interferometer proceeds as shown in Fig. 6.25. Different regions of potential energy are placed behind slit 1 and behind slit 2. A difference in the two potential energies produces a phase shift between the two wave functions that interfere at the distant screen. Hence, a measurement of the fringe shift in the interference pattern is a measurement of the potential energy difference. The different regions might, for example, have different electric fields, which produce different energies in atomic states (see Section 10.7.2). Or, if the atom interferometer is oriented vertically (or at an angle) instead of horizontally, then the two paths experience different gravitational potential energies. Recent experiments have been precise enough to test features of Einstein’s general theory of relativity. Atom interferometers can also measure rotation and acceleration, similar to fiber optic gyroscopes that are commonly used for navigation.''
\end{quotation}
This passage is directly related to sensing, explicitly addresses a quantum sensing platform (atom interferometer), and provides an explanation of the principles behind its operation, thus earning an excerpt sensing score of 6 out of 7.
This excerpt follows a discussion of the double-slit experiment, and it describes how the double-slit technique can be used to measure potential energies and the role of the de Broglie wavelength in enabling this process.
Additionally, it highlights real-world applications of this sensing method (for example, rotation and acceleration).

Another example is a similar excerpt on atom interferometry from Serway (Chapter 10 - Statistical Physics (post-chapter essay), pg 368), tagged under multiple keywords - interference, interferometer, superposition, and sensor:
\begin{quotation}
    ``Atomic fountains have also been used to construct atom interferometers. Similar to an optical interferometer, the atom interferometer splits the atom into a superposition of two coherent states that separate spatially and recombine to form interference fringes. Atom interferometers make extremely sensitive inertial sensors because of the long transit time of the atoms through the device. The Stanford group has measured g, the acceleration of an atom due to gravity, with a resolution of one part in $10^8$ with an atom interferometer, and it is likely that the uncertainty will be reduced to less than one part in $10^{11}$. A portable version of this device could replace the mechanical “g” meters now used in oil exploration. A low value of g could signify the presence of porous, oil-laden rock, which has a lower density than solid rock.''
\end{quotation}
The excerpt includes a description of how and why atom interferometers work as gravity sensors.
It addresses experiments that have been conducted to measure $g$ as well as an application of this idea in a real-life scenario (oil exploration).
Since it is directly relevant to sensing, addresses a quantum sensing platform, and includes a brief conceptual discussion of the sensing protocol, it earns an excerpt sensing score of 5 out of 7.

An example of an excerpt with low sensing score is this excerpt from Griffiths, appearing in the section on the Aharonov-Bohm effect (Chapter 4 - Quantum Mechanics in Three Dimensions, pg 234), tagged under the keyword interference: 
\begin{quotation}
    ``..The beams arrive out of phase by an amount proportional to the magnetic flux their paths encircle ... \textit{This phase shift leads to measurable interference}, which has been confirmed experimentally by Chambers and others''
\end{quotation}
This excerpt does not mention any sensing platform or hardware beyond the existence of a beam of charged particles and lacks a conceptual explanation of the sensing protocol.
However, it references how interference can be used to detect an electromagnetic field by measuring the phase shift it induces, making it slightly relevant to sensing and hence earning 1 point for the Relevance criterion.
As a result, it receives an excerpt sensing score of 1 out of 7.

Next, we will discuss some examples of excerpts that received high, medium, and low depth scores.
Depth can stem from both mathematical and conceptual richness, or solely from conceptual discussion without any mathematical formalism.
If an excerpt lacks mathematical content, its total depth score is likely to be lower despite strong conceptual explanations.
An example of an excerpt with both high mathematical and conceptual depth is the unit on magnetic resonance in McIntyre, tagged under the keyword resonance (Chapter 3 - Schrödinger Time Evolution, pg 87).
This subsection is part of the section on Time-dependent Hamiltonians (section 3.4) and spans five pages, providing a detailed mathematical derivation (equation frequency count score - 3/3)  of the generalized Rabi formula and Rabi flopping.
It begins by defining magnetic resonance, offering a conceptual discussion on the role of Larmor frequency and a comparison between classical and quantum cases of magnetic moment precession, including an analysis of the applied field.
In the quantum mechanical case, the Hamiltonian is defined, followed by a step-by-step solution of the time-dependent Schrödinger equation (effectiveness of math score - 3/3).
The discussion then transitions to the rotating frame, where the new Hamiltonian is derived.
These foundations lead to the calculation of spin-flip probabilities, culminating in an analysis of spin precession and Rabi flopping.
The excerpt includes two figures (diagrams score - 1/1), multiple equations in quantum notation (quantum notation score - 2/2), a rich conceptual explanation (keyword-specific conceptual discussion score - 2/3), and practical applications such as MRI (application score - 1/2), earning it a very high excerpt depth score of 12/14.

An example of an excerpt with high conceptual depth but no mathematical formalism is the discussion tagged under the keyword Entangle— in Mermin, on the bit commitment protocol, which is a cryptographic technique used to commit to a value while keeping it hidden from others. (Chapter 6 - Protocols that Use Just a Few Qubits, pg. 145).
This excerpt falls within the section on bit commitment (section 6.3) , and contains no mathematical content but provides a detailed 1-2 page conceptual explanation of how entangled qubits can be used to cheat in the bit commitment protocol.
It walks through the process step by step, showing how by keeping entangled qubits and choosing her measurement strategy later, Alice can manipulate Bob into believing either outcome of a binary decision.
Additionally, it discusses the broader implications of entanglement for the security of such protocols and quantum cryptography in general.
It does not include any figures or references to direct applications.
These factors lead to a high score of 3/3 for depth of keyword-specific conceptual discussion, while it receives a zero for all other depth criteria, resulting in a low overall excerpt depth score.

To illustrate what a medium depth score represents, we will look at an excerpt from Griffiths (Chapter 1 - The Wave Function, pg. 17) for the keyword measurement.
This passage, part of the section The Statistical Interpretation (Section 1.2), spans roughly two pages and includes a worked example.
It discusses the meaning of the wave function and its interpretation using the idea of quantum indeterminacy.
The excerpt contains one equation, representing the spatial probability of a particle as the integral of the wave function’s amplitude squared (equation frequency score - 1/3, quantum notation score - 2/2).
While this equation relates to the probability of locating a particle during measurement, the mathematical formalism does not fully address the broader conceptual discussion on measurement, resulting in a math effectiveness score of 1/3.
Conceptually, the excerpt explains the particle’s behavior before and after measurement and outlines ``realist, orthodox, and agnostic interpretations" of the nature of the physical system and its properties.
It introduces quantum indeterminacy, emphasizing that a particle does not have a precise position before measurement, and then discusses the implications of repeated (consequent) measurements.
Additionally, it presents a conceptual worked example of the electron interference experiment using double slits (keyword-specific conceptual discussion score - 2/3 for the keyword measurement).
The excerpt includes three figures relevant to measurement : two graphs illustrating the wave function and wave function collapse, and an image of the electron interference pattern (diagrams score: 1/1).
However, it does not mention any applications of the keyword (applications score: 0/2).
Overall, the excerpt earns a total excerpt depth score of 7/14.

An excerpt with a low excerpt depth score would be this passage about relativity experiments in Serway (Chapter 1 - Relativity I, pg. 17).
This excerpt, related to the keyword atomic clock, appears in the sub-section on Time Dilation within the section on Consequences of Special Relativity (Section 1.5).
\begin{quotation}
    ``It is quite interesting that \textit{time dilation can be observed directly by comparing high-precision atomic clocks}, one carried aboard a jet, the other remaining in a laboratory on Earth. \textit{The actual experiment involved the use of very stable cesium beam atomic clocks}. Time intervals measured with four such clocks in jet flight were compared with time intervals measured by reference atomic clocks located at the U.S. Naval Observatory. To compare these results with the theory, many factors had to be considered, including periods of acceleration and deceleration relative to the Earth, variations in direction of travel, and the weaker gravitational field experienced by the flying clocks compared with the Earth-based clocks.''
\end{quotation}
This excerpt contains no mathematical content related to atomic clocks (equation frequency, effectiveness, and quantum notation usage scores – 0) and lacks any relevant diagrams (diagrams score – 0).
It briefly discusses how atomic clocks are used in experiments to observe time dilation, describing the experimental setup and comparison of experimental results with the theoretical predictions.
However, it treats atomic clocks solely as tools for the experiment without explaining their underlying principles or the process of time measurement, resulting in a keyword-specific conceptual discussion depth score of 1/3.
Since the excerpt presents an application of atomic clocks, it earns an application score of 1/2, leading to a total excerpt depth score of 2/14.

\subsection{Network analysis}\label{Res:network}

\begin{figure*}[p] 
    \centering
    \includegraphics[width=\textwidth,height=\textheight,keepaspectratio]{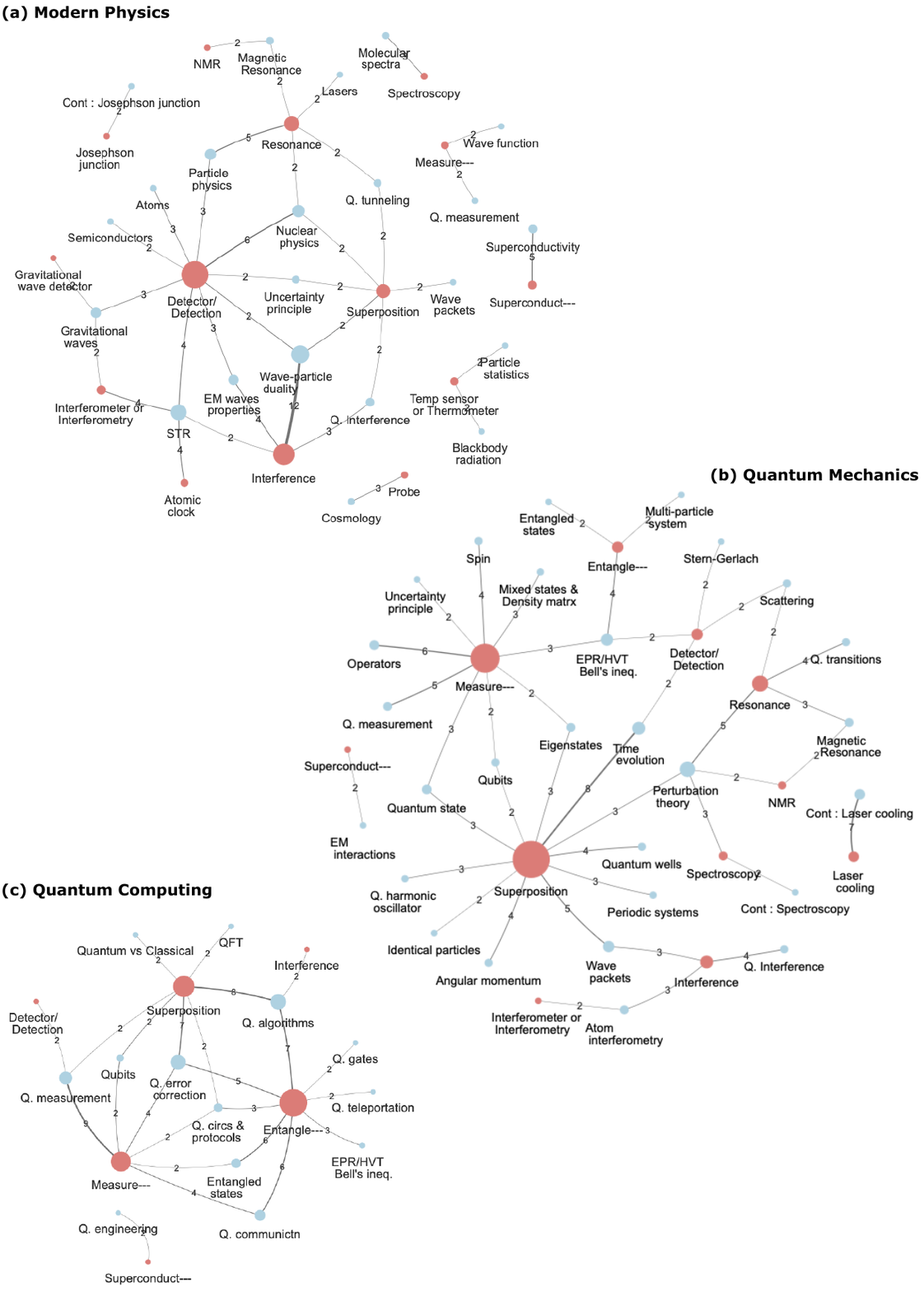}
    \caption{Network maps showing the contexts in which the keywords appear. Each diagram includes data from the two textbooks for that subject. The red nodes are the keywords and the blue nodes are the contexts. Abbreviations : STR - Special Theory of Relativity, EPR - Einstein–Podolsky–Rosen experiment, HVT - Hidden variable theory, NMR - Nuclear Magnetic Resonance, QFT - Quantum Fourier Transform}
    \label{fig:networkmaps}
\end{figure*}

To visualize the contexts in which different keywords appear across textbooks, we constructed three separate network diagrams, each corresponding to data from the two textbooks analyzed in a given subject (Fig. \ref{fig:networkmaps}).
These diagrams represent the relationships between keywords and their respective contexts, providing insight into how frequently different concepts are associated within the textbooks.

In each diagram, keywords are represented as red nodes, while contexts in which they appear are shown as blue nodes.
An edge is drawn between a keyword and a context whenever an excerpt for that keyword appears in that specific context.
The thickness of the edge, along with the number displayed on it, indicates how many excerpts for that keyword appeared in that context.
The size of each node represents the total number of occurrences (number of excerpts) of that keyword or context in the data.

For better clarity of the visualizations, edges representing only a single occurrence (edge width of 1) are omitted from the diagrams.
Consequently, nodes without any remaining connections, due to the removal of single-occurrence edges, are also excluded.
As a result, certain keywords may have appeared in many more excerpts than what is directly visible in the figure, particularly if those occurrences were distributed across multiple contexts but only appeared once in each.
Therefore, the visualization in Fig. \ref{fig:networkmaps} should not be interpreted as a direct indicator of the total number of excerpts for the keyword within that subject.
We have included the complete version of these network maps in the Supplemental Material \cite{supp}, where we include all the omitted weight-1 edges, to show that the general trends we observe here hold true.

We have also constructed two additional network maps that compare the spin-first and position-first textbooks, which we will discuss in detail in Sec.\ \ref{Res:pos_vs_spin}.
In the following sections, we will analyze the heatmaps, discuss the sensing and depth scores, and discuss the network analysis results for each category of keywords listed in Table \ref{tab:keywords}.

\subsection{Heatmap : Core concepts}\label{Res:sec:heatmap_coreconcepts}

\begin{figure}[H]
\includegraphics[scale=0.97]{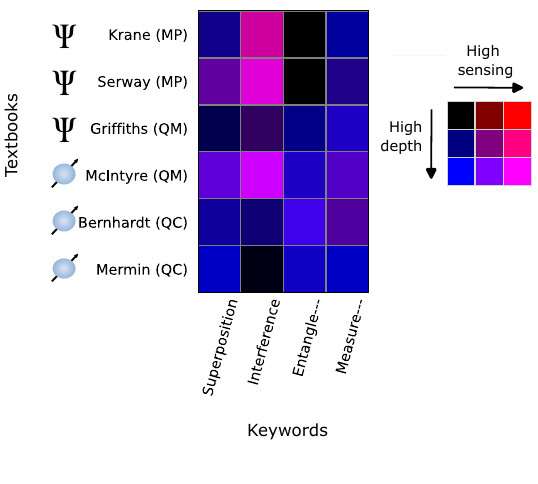}
\caption{\label{fig:hm_phenomena}Heatmap showing the sensing and depth scores for the keyword category - `Core concepts'. The icons next to the textbook names indicate whether they use a position-first (\includegraphics[height=1.8ex]{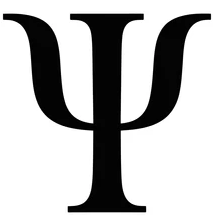}) or spin-first (\includegraphics[height=1.8ex]{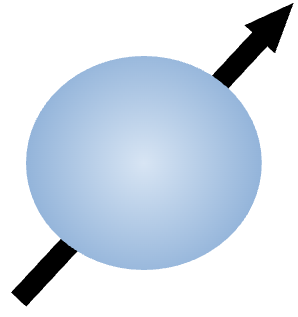}) approach. The color mixing scheme used in the heatmap is described in section \ref{subsec:Method_rubric_heatmap}. Although the legend shows a discrete set of colors, the heatmap itself uses a continuous color scale.}
\end{figure}

We consider the keywords superposition, interference, entanglement, and measurement as core concepts because of the central role they play in quantum sensing. From the definition presented earlier, quantum sensing uses the quantum phenomena of superposition, interference, and entanglement to achieve a quantum advantage, while measurement is an essential step in any sensing protocol. As a result, most quantum sensing concepts or protocols are built on one or more of these fundamental ideas.

In this section, we analyze the heatmap (Fig. \ref{fig:hm_phenomena}) for keywords in the `core concepts' category (see the keyword categories in Table \ref{tab:keywords}).
We start with high depth and high sensing score examples, focusing on the keyword interference.
We then move on to examples of high depth but low sensing - starting with the keyword superposition, then entanglement, and finally measurement.

\textbf{Interference} receives high sensing and high depth scores in both the modern physics textbooks as well as McIntyre's \textit{Quantum Mechanics} (Fig. \ref{fig:hm_phenomena}).
As the Modern Physics network map in Fig. \ref{fig:networkmaps}(a) shows, one significant context for interference is special theory of relativity (STR), where the Michelson–Morley experiment and its implementation to sense the presence of the luminiferous ether is explained in detail in both Krane and Serway.
In Serway, this section provides a detailed three-page mathematical derivation of the protocol, includes diagrams, and offers a conceptual explanation of how interference patterns were used to try to detect the presence of ether.

In Krane, interference also appears context of Josephson junctions, where their application in superconducting quantum interference devices (SQUIDs), and their use in magnetic field sensing are addressed within the broader context of superconductivity.
In Krane, the highest depth excerpt is a different section on ‘Interference and Diffraction’, presented in the context of electromagnetic waves.
This three-page section discusses the classical perspective of optical interference using Young’s double-slit experiment, provides the mathematical derivation for locating maxima and minima, incorporates diagrams, and offers a conceptual discussion of the physics behind interference and diffraction.

In Serway, the use of atomic fountains in atom interferometry for sensing purposes is introduced in a post-chapter essay on laser manipulation of atoms.
While the atom interferometry sensing protocol itself is not discussed in depth, the section offers a brief insight into this quantum sensing platform, making it a high sensing but low depth excerpt.
Another example of a high depth excerpt in Serway is the description of electron diffraction in terms of $\psi$, situated in the context of wave–particle duality and quantum interference (as seen in Fig. \ref{fig:networkmaps}(a)).
This excerpt includes significant mathematical detail, contains a rich conceptual discussion of the double-slit experiment, the wave–particle duality of electrons, and an explanation of probability in terms of the wavefunction, spanning approximately five pages.

The high sensing and highest depth excerpt in McIntyre is a dedicated section on atom interferometry (as seen in Fig. \ref{fig:networkmaps}(b)), explaining the protocol for potential energy measurements with supporting diagrams and applications, giving it a high sensing score.
This section offers a two-page explanation of the optical double-slit experiment with full mathematical details, followed by a discussion of the double-slit experiment with particles to illustrate quantum interference, and finally a detailed explanation of the atom interferometry protocol for potential energy measurements.
This last part provides a rich conceptual discussion, though it uses only minimal mathematical formalism.
In McIntyre, interference also appears in the context of quantum interference, notably in discussions of the Stern–Gerlach experiment, and of wave packets, in discussions about free particle states and discrete and continuous superpositions of momentum eigenstates (Fig. \ref{fig:networkmaps}(b)).

The keyword interference has very few excerpts and receives very low scores in Griffiths' \textit{Quantum Mechanics} and the quantum computing books (Fig. \ref{fig:hm_phenomena}).
In Griffiths, interference appears mainly in contexts such as quantum interference, particularly in reference to the wavefunction, wave packets, and electron interference experiments (Fig. \ref{fig:networkmaps}(b)).
Even though these contexts are similar to those in McIntyre, Griffiths lacks the in-depth sensing related discussions present in McIntyre.
In the quantum computing textbooks (Fig. \ref{fig:networkmaps}(c)), interference appears repeatedly only in the context of quantum algorithms, where it is used to qualitatively describe the interference of terms within a superposition state and how it can be leveraged for the algorithms, such as the Shor's algorithm in Bernhardt and the Bernstein–Vazirani problem in Mermin.

\textbf{Superposition - }Moving on to keywords that are discussed in depth but not within sensing-related contexts, the keyword superposition receives high depth scores in both McIntyre's \textit{Quantum Mechanics} and Mermin's quantum computing books (Fig. \ref{fig:hm_phenomena}).
None of the textbooks had high sensing excerpts for superposition.
In McIntyre, excerpts generally receive low sensing scores, although a few discussions have a slight relevance to sensing - for example, Rabi oscillations, magnetic resonance (context - magnetic resonance), and the Stark effect (context - perturbation theory, Fig. \ref{fig:networkmaps}(b)) (all were low depth excerpts).
However, McIntyre contains multiple high depth excerpts about superposition. One example is the section on superposition states in an infinite square well.
This excerpt spans nearly five pages and includes a fully worked example with multiple figures.
It provides detailed mathematical derivations concerning the time evolution of superposition states, the probability of energy measurements, the expectation value of energy, and the calculation of expansion coefficients.
Other high depth excerpts include discussions on the superposition state of a particle on a ring in the context of angular momentum and superposition states in the context of qubits.
Additional contexts where superposition appears in McIntyre include eigenstates, the quantum harmonic oscillator, quantum states (state vectors and superposition states), wave packets, periodic systems (molecular states, periodic wells, semiconductors), and perturbation theory (hyperfine Hamiltonian and the Stark effect) (Fig. \ref{fig:networkmaps}(b)).

In Mermin, an example of a high depth excerpt is the section on measurement gates and the Born rule (context - quantum measurement, Fig. \ref{fig:networkmaps}(c)).
This six-page section presents a rich conceptual discussion of the measurement of n-qubit states written as superpositions of computational basis states.
It also covers measurement gates, state collapse, and the effects of Hadamard and projection operators.
While this section is not mathematically dense, it provides sufficient formalism to illustrate key concepts, supported by mathematical examples.
Other contexts in which superposition is discussed in Mermin include quantum algorithms, quantum error correction, QFT, and quantum circuits and protocols (Fig. \ref{fig:networkmaps}(c)).

In contrast, Griffiths' QM and the modern physics textbooks contain only a few excerpts tagged under superposition.
Griffiths receives a low depth score (Fig. \ref{fig:hm_phenomena}), with the few relevant excerpts focused mainly on wave packets and with limited discussion of quantum superposition itself.
This is an interesting difference between Griffiths, a position-first textbook, and McIntyre, which follows a spin-first approach and has a more explicit connection to quantum information science and engineering.

The modern physics textbooks receive a medium depth score (in the range 7-8 out of 14)(Fig. \ref{fig:hm_phenomena}).
While they do not emphasize quantum superposition in depth, excerpts touch on a broad range of contexts, including the uncertainty principle, wave-particle duality (the electron diffraction experiment), quantum interference, tunneling, wave packets, and nuclear physics (superposition of electromagnetic fields) (Fig. \ref{fig:networkmaps}(a)).
Serway briefly mentions the role of atomic superposition states in atom interferometry, resulting in a slightly higher (but still low) sensing score (Fig. \ref{fig:hm_phenomena}).

Bernhardt, the other quantum computing book, contains some key discussions on superposition in the contexts of quantum algorithms, quantum cryptography (single occurrence), quantum measurement, etc (Fig. \ref{fig:networkmaps}(c)) and receives a medium depth score (Fig. \ref{fig:hm_phenomena}).

\textbf{Entanglement - }The keyword entangle- receives high depth scores in Bernhardt (12/14), and slightly lower but still very high depth scores in Mermin (10/14) and McIntyre (10/14), represented by the dark blue cells on the heatmap - Fig. \ref{fig:hm_phenomena}.
These discussions are important as they help build foundational understanding of key concepts essential for learning about quantum sensing.
In Bernhardt, there are no excerpts directly related to sensing, except for a brief discussion on the properties of photons (for example, easy to maintain entanglement over long distances due to reduced interactions with the environment) that hold some relevance for quantum technology, particularly hardware platforms.
However, the textbook contains several high depth excerpts, such as the sections on superdense coding and quantum teleportation, which provide complete step-by-step protocols, including circuit diagrams and mathematical descriptions.
These excerpts, situated within the context of quantum communication (Fig. \ref{fig:networkmaps}(c)), highlight the use of entanglement as a central tool in these protocols.
Similarly, Mermin does not include any sensing-relevant discussions, but its treatments of Bell state preparation and teleportation are rich in both mathematical and conceptual content, both including step-by-step protocols and circuit diagrams.
In the quantum computing textbooks, entanglement is also discussed in the contexts of the theory of entangled states, quantum algorithms, quantum error correction, the EPR paradox/hidden variable theory, and quantum gates (Fig. \ref{fig:networkmaps}(c)).

In McIntyre, a similar section on quantum teleportation (in the final chapter on Modern Applications of Quantum Mechanics, which mostly covers QISE-related topics) provides a complete mathematical description of the protocol, making it a high depth excerpt for ‘entangle—’.
Entanglement is also addressed in discussions of the EPR paradox/hidden variable theory (Fig. \ref{fig:networkmaps}(b); abbreviated as EPR/HVT/Bell's ineq.) and the theory of entangled states, where the book introduces entangled states and their properties which can be leveraged in QISE.
Entanglement does not appear at all in the modern physics textbooks and has only a few low-depth occurrences in Griffiths, with very brief conceptual and mathematical discussions of the concept.
It is mentioned primarily in the context of the EPR paradox and multi-particle systems in Griffiths.

\textbf{Measurement - }Another example of a high depth but low sensing score is the keyword measure—, as it receives high depth scores in Griffiths, McIntyre, and Mermin (Fig. \ref{fig:hm_phenomena}).
Bernhardt receives a medium depth score but still includes several meaningful discussions.
Our analysis considers only those instances where the excerpt explicitly addresses the quantum aspect of measurement.
As a result, although modern physics textbooks contain numerous mentions of the word, these were not included unless they directly referred to quantum measurement, and thus do not contribute to the score.

In Griffiths, section on Generalized Statistical Interpretation (contexts - quantum interference and quantum measurement provides an in-depth treatment of quantum measurement - Fig. \ref{fig:networkmaps}(b)).
This three-page discussion covers the statistical interpretation of the wavefunction, probability amplitudes, and observables, including detailed mathematical derivations of the expectation value, measurement probabilities, and the Fourier transform between position and momentum spaces.
High depth excerpts in McIntyre include the discussion on the EPR experiment, hidden variable theory, and Bell’s inequality.
This five-page discussion explains how EPR used the concept of measurement to argue about elements of reality, locality, and hidden variables in quantum mechanics.
It presents a mathematical description of the EPR paradox, a detailed conceptual explanation of the role of measurement in understanding the ``objective reality" of the properties of physical systems, and a discussion on understanding Bell’s inequality using measurement probabilities, also including figures and tables.
Other contexts in which measurement appears in the quantum mechanics textbooks (Fig. \ref{fig:networkmaps}(b)) include operators (Hermitian operators, observables, and determinate states), spin (spin-1/2 systems, electrons in a magnetic field, and measurement of spin components along different directions), mixed states and density matrix, and quantum states (stationary and superposition states).

In Mermin, for the keyword measure-, multiple high depth excerpts are present.
Topics such as measurement gates, the Born rule, and probability amplitudes are the main topics in the quantum measurement-related content (Fig. \ref{fig:networkmaps}(c)).
For instance, the section on the Generalized Born Rule (context - quantum measurement, Fig. \ref{fig:networkmaps}(c)) provides a detailed two-page mathematical explanation of measuring one of many qubits, also including circuit diagrams.
It also addresses orthogonality and normalization of states, and what it means in reference to measurements.
The discussion on the Quantum Fourier Transform (QFT) is another example of an excerpt with detailed explanation on the role of measurement gates (context - QFT, but not shown in Fig. \ref{fig:networkmaps}(c) due to single occurrence).
Similarly, Bernhardt contains rich discussions on the Born rule, measurement gates, and measurement along different bases.
In the quantum computing textbooks, measurement is also explored in the contexts of quantum error correction (error diagnosing syndromes and correcting codes) and quantum communication (particularly cryptography, impossibility of superluminal communication, etc) (Fig. \ref{fig:networkmaps}(c)).

The keyword `measure-' receives medium depth scores in the modern physics textbooks, as references to quantum measurement are limited (Fig. \ref{fig:hm_phenomena}).
When present, they appear mainly in the context of wavefunctions (probability density) and the theory of quantum measurement, including probability, randomness, and expectation values (Fig. \ref{fig:networkmaps}(a)).
In Serway, a section on `The Born Interpretation' introduces the idea of Born rule in the context of wavefunctions.

\subsection{Heatmap : Techniques / Applications}\label{Res:sec:heatmap_applicns}

\begin{figure*}[htbp]
\includegraphics[scale=0.9]{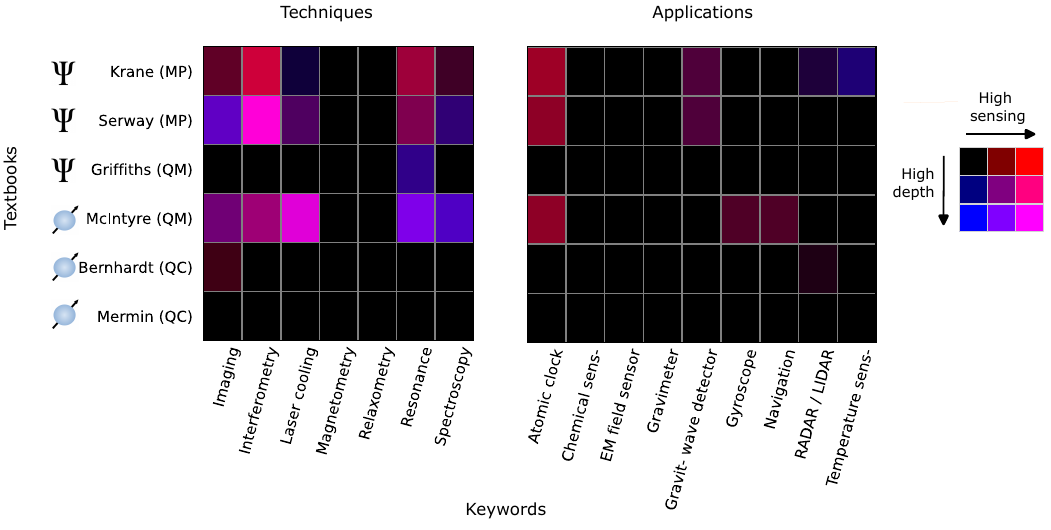}
\caption{\label{fig:hm_applicns}Heatmap showing the sensing and depth scores for the keyword category - `Techniques / Applications'. The icons next to the textbook names indicate whether they use a position-first (\includegraphics[height=1.8ex]{psi_figure.png}) or spin-first (\includegraphics[height=1.8ex]{spin_figure.png}) approach. The color mixing scheme used in the heatmap is described in section \ref{subsec:Method_rubric_heatmap}. Although the legend shows a discrete set of colors, the heatmap itself uses a continuous color scale.}
\end{figure*}

In this section, we analyze the heatmap for keywords in the `techniques / applications' category (Fig. \ref{fig:hm_applicns}).
We start with high depth and high sensing score examples like the keywords interferometry and laser cooling.
We then focus on the high sensing but low depth examples for the keywords atomic clock and resonance.
We then move on to high depth and low sensing examples, analyzing the keywords imaging and spectroscopy.
Finally, we briefly discuss some of the other keywords in the applications sub-category.

\textbf{Interferometry /Interferometer} receives high sensing and depth scores in Serway’s \textit{Modern Physics} (Fig. \ref{fig:hm_applicns}).
It is worth noting that interferometry is a technique and interferometer is a device, which we distinguish from interference, which is a core concept or phenomenon. 
But as this keyword is closely related to interference, many excerpts are common between the two keywords.
For example, similar to interference, the excerpt on the Michelson-Morley experiment in Serway, discussed in the context of special relativity (Fig. \ref{fig:networkmaps}(a)), receives a high excerpt sensing score.
It also receives a high excerpt depth score, as it explains the full interferometry protocol (where the goal is to detect the change in relative velocity between apparatus and ether by measuring the shift of interference fringes) with a detailed mathematical derivation, and the keyword gets a high keyword depth score.
Other excerpts in Serway with high sensing scores include a paragraph on atom interferometry and a brief description of gravitational wave detection using LIGO.
These excerpts, though informative in terms of sensing protocol and applications, are only one or two paragraphs long and lack deep mathematical or conceptual discussions on interferometry, resulting in low excerpt depth scores.
The keyword interferometry receives a high sensing but low depth score in Krane (Fig. \ref{fig:hm_applicns}).
A similar excerpt on the Michelson-Morley experiment receives a high excerpt sensing score, and another high-sensing excerpt on gravitational wave detection further contributes to the overall high keyword sensing score.
However, none of these excerpts contain detailed mathematical or conceptual discussions, leading to a low keyword depth score in Krane.

In quantum mechanics, interferometry does not appear at all in Griffiths but has a few excerpts in McIntyre.
It should be noted that the keyword interference does appear in Griffiths in discussions of electron interference and the double-slit experiment.
However, for the purposes of this analysis, we do not classify this as interferometry, as we are trying to distinguish between the concept of interference and its application within an apparatus to achieve a specific task.
One excerpt in McIntyre explains the protocol for atom interferometry for measuring gravitational potential energy, earning a high excerpt sensing score.
However, due to the small number of excerpts overall, the keyword receives only a medium sensing score.
Since the atom interferometry excerpt lacks mathematical depth, it gets a medium excerpt (and keyword) depth score (Fig. \ref{fig:hm_applicns}).
Another context in which interferometry appears in McIntyre is the Stern-Gerlach experiment, but this is not shown in the network diagram (Fig. \ref{fig:networkmaps}(b)) as this connection (edge) occurs only once.
In the quantum computing textbooks, interferometry as a technique does not appear at all.

\textbf{Laser cooling} is another example of a keyword with high sensing and depth scores in McIntyre’s \textit{Quantum Mechanics} (Fig. \ref{fig:hm_applicns}).
McIntyre includes an entire subsection on laser cooling within the section on manipulating atoms with quantum mechanical forces.
This section spans nearly eight pages, explains the basic principles behind laser cooling and the laser cooling cycle with diagrams and equations, discusses the Doppler effect, and provides a mathematical derivation of the scattering force.
It also covers advanced topics such as chirped cooling and optical molasses, resulting in a very high excerpt (and thus keyword) depth score.
Laser cooling is highly relevant to atom-based sensing platforms, and this excerpt explicitly discusses its applications in quantum sensing, including atomic clocks and gravity measurements, hence receiving a high excerpt sensing score.
Laser cooling is also mentioned briefly in the context of atom interferometry, which contributes to the high keyword sensing score.
In McIntyre, although the keyword mostly appears in the context of laser cooling itself (Fig. \ref{fig:networkmaps}(b)), it also appears once in the context of magnetic trapping.
The keyword laser cooling does not appear at all in Griffiths.

In modern physics, Krane briefly mentions laser cooling under applications of Bose-Einstein condensates in the context of particle statistics, but this excerpt (and hence the keyword) receives very low sensing and depth scores (Fig. \ref{fig:hm_applicns}).
In Serway, a post-chapter essay on laser manipulation of atoms includes a section on laser cooling.
This section contains a rich conceptual discussion on the theory of laser cooling, optical molasses, Raman cooling, and related topics, including diagrams and references to applications.
However, due to the lack of mathematical depth, it receives a low-medium excerpt (and thus keyword) depth score (Fig. \ref{fig:hm_applicns}).
This excerpt addresses the application of atomic fountains in atom interferometry and briefly explains the working of an atom interferometer, which gives it a high excerpt sensing score.
But the small number of excerpts lead to an overall low-medium keyword sensing score (Fig. \ref{fig:hm_applicns}).
Laser cooling does not appear at all in either quantum computing textbooks.

We will now discuss examples of some keywords that receive high sensing, but low depth scores.

\textbf{Atomic clock - }The keyword atomic clock receives high sensing and low depth scores in both the modern physics books and McIntyre’s \textit{Quantum Mechanics} (Fig. \ref{fig:hm_applicns}).
But it has only minimal coverage as there are only 2-3 short excerpts in each of these textbooks.
In Krane’s \textit{Modern Physics}, an atomic clock is described as a tool used for experimental verification of time dilation and twin paradox in the context of special relativity (Fig. \ref{fig:networkmaps}(a)).
These two excerpts receive high sensing scores due to their direct relevance to sensing and explicit mention of a sensing platform and its application.
In Serway, in addition to a similar excerpt on special relativity experiments, atomic clocks are also mentioned in the context of laser cooling in the post-chapter essay on laser manipulation of atoms.
These excerpts contribute to the high keyword sensing score, as they directly address the sensing applications of atomic clocks.
But none of these excerpts in Krane or Serway have any detailed conceptual or mathematical discussion on the working and usage of atomic clocks, and hence receive low depth scores.

In \textit{Quantum Mechanics} by McIntyre, the high sensing excerpt on atomic clocks involves a brief discussion on how hyperfine transitions can be used for precise frequency measurements, in the context of `energy spectrum of atoms'.
Another excerpt on the application of laser cooling also mentions atomic clocks and its relevance to sensing, which further contributes to the high keyword sensing score in McIntyre.
But these excerpts also lack mathematical and conceptual details in their discussion of atomic clocks and receive low depth scores.
Atomic clocks are not mentioned in Griffiths' \textit{Quantum Mechanics} and the quantum computing textbooks. 

\textbf{Resonance} is another keyword with high sensing but low depth scores in Krane’s \textit{Modern Physics} (Fig. \ref{fig:hm_applicns}).
In this textbook, the high sensing excerpt appears in the context of superconductivity in a discussion on the application of Superconducting Quantum Interference Devices (SQUIDs) in Magnetic Resonance Imaging (MRI).
This excerpt addresses the magnetic field sensing protocol used in a SQUID, and hence receives a high sensing score.
However, it does not contain any rich discussions on resonance or MRI, which gives it a low depth score.
The keyword resonance also appears in the context of particle physics (Fig. \ref{fig:networkmaps}(a)), but here the word resonance is used to describe resonance particles, which are extremely short-lived states, the existence of which can be indirectly inferred from its decay products.
Due to its relevance in particle detection, this excerpt also contributes to the high keyword sensing score, but the lack of rich conceptual and mathematical discussion on resonance gives it a low depth score.
Serway’s \textit{Modern Physics} has more excerpts for resonance but all are low sensing and low depth excerpts.
Resonance also appears in contexts such as lasers and nuclear physics, in both Krane and Serway.
Additionally in Serway, it appears in the contexts of magnetic resonance, quantum tunneling, and laser cooling (some of which are not shown in (Fig. \ref{fig:networkmaps}(a)) because it occurs only once).
It is interesting to note that Krane does not contain any discussions of magnetic resonance or NMR, other than passing mentions of MRI in a section on superconductivity. 

In quantum mechanics, the keyword resonance receives a medium depth score in Griffiths and a high depth score in McIntyre, but receives low sensing scores in both the textbooks (Fig. \ref{fig:hm_applicns}).
Griffiths has only a few excerpts for resonance, such as an excerpt in the context of quantum transitions focused on Fermi’s golden rule and the resonance condition.
This excerpt uses one equation, a diagram, and a brief conceptual discussion to address the resonance condition, but does not involve any sensing-related discussion, which earns it a medium depth score and low sensing score.
Another example of a medium depth excerpt is a problem on magnetic resonance in a spin-half system, which includes some mathematical details.
McIntyre, on the other hand has many excerpts, with one high depth excerpt, which is a subsection on magnetic resonance within the section on time-dependent Hamiltonians.
This 5 page-long excerpt includes a detailed mathematical derivation of the Rabi formula, and rich conceptual discussions on Larmor precession, magnetic resonance, and Rabi oscillations, supported by diagrams.
However, this excerpt does not address any sensing-related topics, and hence receives a low sensing score.
Scattering, perturbation theory, laser cooling, and spectroscopy are some contexts in which resonance appears in McIntyre, apart from magnetic resonance (Fig. \ref{fig:networkmaps}(b)).
Resonance does not receive any mention in the quantum computing textbooks. 

\textbf{Imaging} is a keyword with high depth but low sensing scores in Serway’s \textit{Modern Physics} (Fig. \ref{fig:hm_applicns}).
Although imaging has minimal coverage in modern physics textbooks, Serway includes a post-chapter essay on the scanning tunneling microscope (STM) in the context of quantum tunneling, which contains a comprehensive discussion on the imaging technique.
The section covers the theory behind the operation of an STM and includes multiple diagrams and a few equations to explain concepts such as the work function, characteristic length scale, and tunneling current density.
As a result, this excerpt receives a very high excerpt (and keyword) depth score.
However, most of the excerpts in Serway do not address sensing, resulting in a low overall keyword sensing score.
The only exception is a subsection on Nuclear Magnetic Resonance and Magnetic Resonance Imaging (MRI) in the chapter on nuclear structure, which briefly explains the imaging technique used in MRI and receives a medium excerpt sensing score.
In Krane, there are only two excerpts tagged under imaging.
One is a subsection on SQUIDs in the context of superconductivity, which describes the magnetic field sensing protocol and mentions applications in MRI.
This excerpt receives a medium-high sensing score but, since it does not explore imaging in detail, it gets a low excerpt depth score.
Other contexts in which imaging appears in modern physics include gravitational waves, cosmology, and particle physics, although these instances are not shown in the network diagram as they occur only once in each context.

In quantum mechanics, imaging does not appear at all in Griffiths, but McIntyre includes two excerpts.
One excerpt is on the STM, which explains how it works, how tunneling is leveraged in the device, and includes a couple of equations related to tunneling probability and current.
This excerpt contributes to the medium keyword sensing and keyword depth scores.
Beyond this context of quantum tunneling, imaging appears once more in the context of magnetic resonance.
Within the quantum computing textbooks, the keyword imaging appears only once in Bernhardt, as a very brief mention in the context of quantum engineering or hardware.

\textbf{Spectroscopy} is another example, with high depth but low sensing scores in McIntyre’s \textit{Quantum Mechanics} (Fig. \ref{fig:hm_applicns}).
The high-depth excerpt is a 2-page long section dedicated to spectroscopy, and it provides a detailed conceptual and mathematical description of the absorption and emission spectra, energy levels, and a typical spectroscopy experiment, including energy level diagrams and tables of resonant frequencies.
None of the excerpts for this keyword directly address any sensing-related topics, which lead to a low keyword sensing score.
Perturbation theory (Fig. \ref{fig:networkmaps}(b)), quantum transitions, quantum harmonic oscillator, and identical particles are some other examples of the contexts in which spectroscopy appears in McIntyre.
Griffiths contains only two passing mentions of spectroscopy, neither of which appear in the main text.
One of these occurrences is in a problem on angular momentum, and the other is in a footnote in the context of band structure.

Spectroscopy is not covered in depth in Krane’s \textit{Modern Physics}, and it receives very low keyword sensing and depth scores.
Serway also has only a few excerpts, but it does include a detailed conceptual discussion on the theory and applications of spectroscopy in the context of energy spectrum of atoms (the Bohr model of atom), which receives a medium excerpt depth score, but low excerpt sensing score.
Spectroscopy occurs mostly in the context of molecular spectra in the modern physics textbooks (Fig. \ref{fig:networkmaps}(a)), and once in nuclear physics.
The quantum computing textbooks do not address spectroscopy at all. 

\textbf{Other `Applications' keywords - }Among the keywords in the applications subcategory (Fig. \ref{fig:hm_applicns}), most receive low sensing and low depth scores (except atomic clock, which we have already discussed previously).
Gravitational wave detector, gyroscope, navigation, and temperature sensor are some examples.
Gravitational wave detector is a topic addressed to some extent in both modern physics textbooks.
Both books explain the sensing mechanism used in the Laser Interferometer Gravitational-wave Observatory (LIGO), resulting in medium excerpt (and keyword) sensing scores.
However, the lack of mathematical depth and diagrams, and the very brief conceptual explanation in these discussions lead to an overall low depth score.
These discussions appear under sections on general relativity in the context of gravitational waves in both textbooks (Fig. \ref{fig:networkmaps}(a)).
In McIntyre’s \textit{Quantum Mechanics}, the measurement of gravitational potential energies is briefly mentioned in the context of atom interferometry, along with its applications in gyroscopes used for navigation (hence the scores associated with the keywords navigation and gyroscope in Fig. \ref{fig:hm_applicns}).
Krane’s \textit{Modern Physics} includes a couple of excerpts tagged under the keyword temperature sensor / thermometer, both within the context of particle statistics (Fig. \ref{fig:networkmaps}(a)).
One of these is a detailed problem on measuring the temperature of cold systems using the relative population of magnetic substates of nuclei, giving it a low-medium keyword depth score (Fig. \ref{fig:hm_applicns}).
Serway also includes two mentions of this keyword in the context of blackbody radiation (Fig. \ref{fig:networkmaps}(a)), but these receive low sensing and depth scores.

\subsection{Heatmap : Hardware}\label{Res:sec:heatmap_hardware}

\begin{figure*}[htbp]
\includegraphics[scale=0.95]{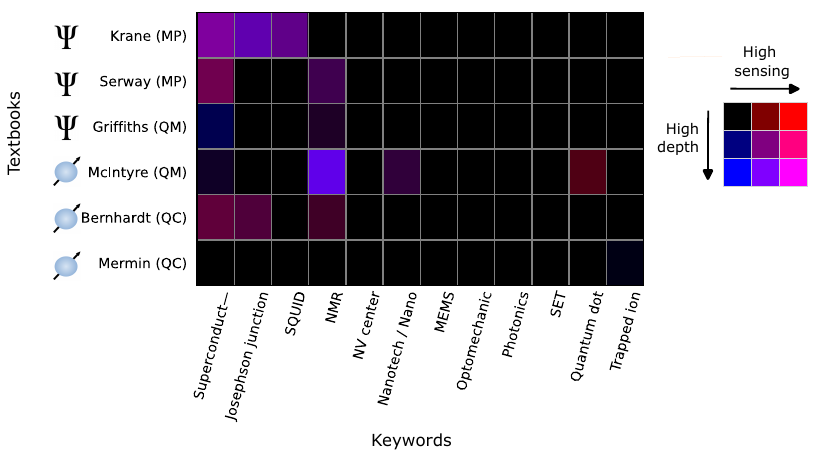}
\caption{\label{fig:hm_hardware}Heatmap showing the sensing and depth scores for the keyword category - `Hardware'. The icons next to the textbook names indicate whether they use a position-first (\includegraphics[height=1.8ex]{psi_figure.png}) or spin-first (\includegraphics[height=1.8ex]{spin_figure.png}) approach. The color mixing scheme used in the heatmap is described in section \ref{subsec:Method_rubric_heatmap}. Although the legend shows a discrete set of colors, the heatmap itself uses a continuous color scale.}
\end{figure*}

In this section, we analyze the heatmap for keywords in the `hardware' category (Fig. \ref{fig:hm_hardware}). No keywords in this category receive both high sensing and depth scores, so we start with discussing the keyword superconduct---, which receives medium sensing and medium depth scores (in Krane). We then discuss high depth and medium sensing examples for the keyword NMR. Finally, we look at the low depth and low sensing examples for the keyword quantum dot.

\textbf{Superconduct---} receives medium sensing and medium depth scores in Krane’s \textit{Modern Physics} (Fig. \ref{fig:hm_hardware}).
There are only four excerpts for this keyword in Krane, one of which is a subsection on the Josephson effect.
This four-paragraph-long discussion, situated in the context of Josephson junctions, explains the theory of the Josephson effect and its application in building a SQUID and magnetic field sensing.
It includes a few equations and figures resulting in a medium excerpt (and keyword) depth score.
It receives a high excerpt sensing score, but the small number of excerpts overall leads to a medium keyword sensing score.
Since this excerpt addresses both Josephson junctions and SQUIDs (each of which are also keywords), these keywords receive similar scores in Krane (Fig. \ref{fig:hm_hardware}).
Another context in which superconduct— appears is a section on the phenomenon of superconductivity (Fig. \ref{fig:networkmaps}(a)), where the excerpt features rich conceptual discussion on the theory and applications of superconductors, BCS theory, high-temperature superconductivity, etc.
However the lack of mathematical depth gives this excerpt a medium depth score and it receives a low excerpt sensing score.
In Serway’s \textit{Modern Physics}, the keyword receives low depth scores and medium sensing scores (Fig. \ref{fig:hm_hardware}).
Similar to Krane, Serway includes a rich conceptual discussion on superconductivity but without substantial mathematical detail.
The keyword earns a medium keyword sensing score because some excerpts include a problem mentioning the use of superconducting magnets in NMR spectroscopy in the context of magnetic resonance, and a discussion on its use in particle tracking and detection (superconducting magnets that produce strong magnetic fields in the accelerator) within the context of particle physics.

In both quantum mechanics textbooks, superconduct— appears only a few times and receives low sensing and depth scores (Fig. \ref{fig:hm_hardware}).
In Griffiths, the keyword is mentioned twice in the context of electromagnetic interactions (Fig. \ref{fig:networkmaps}(b))- specifically within a problem on superconducting rings and flux quantization.
In the quantum computing textbooks, the keyword appears a few times in Bernhardt, where one excerpt receives a high excerpt sensing score for its discussion of Cooper pairs and Josephson junctions in the context of quantum engineering (Fig. \ref{fig:networkmaps}(c)).
The excerpt mentions the use of Josephson junctions to measure magnetic fields only briefly and does not delve into the details of the sensing process or discuss topics like SQUID.
However, the limited number of excerpts results in a medium keyword sensing score, and the lack of depth gives it a very low excerpt (and keyword) depth score (Fig. \ref{fig:hm_hardware}).
Consequently, the keyword Josephson junction also receives similar scores in Bernhardt (Fig. \ref{fig:hm_hardware}).
There is no mention of the keyword superconduct--- in Mermin.

\textbf{NMR - }An example of a keyword with high depth and medium sensing scores is Nuclear Magnetic Resonance (NMR) in McIntyre’s \textit{Quantum Mechanics} (Fig. \ref{fig:hm_hardware}).
Although there are only three excerpts in the textbook tagged under this keyword, one of them is a full subsection on the topic within the section on time-dependent Hamiltonians, in the context of magnetic resonance (Fig. \ref{fig:networkmaps}(b)).
This excerpt presents a detailed theoretical and mathematical description of Larmor precession, Rabi flopping, and applications of NMR, and includes multiple diagrams, which gives it a high excerpt (and therefore keyword) depth score.
The keyword received a medium keyword sensing score due to a few excerpts related to NMR spectroscopy and the role of chemical shifts in identifying chemical microstructures.
In McIntyre, it also appears in the context of perturbation theory (Fig. \ref{fig:networkmaps}(b)).
In Griffiths it appears only once in a problem related to magnetic resonance (Fig. \ref{fig:networkmaps}(b)).

There is no discussion of NMR in Krane’s \textit{Modern Physics}, but Serway includes a subsection within the section titled Some Properties of Nuclei.
This excerpt offers a rich conceptual discussion of NMR, but it receives a low depth score due to the lack of mathematical depth.
It earns a medium excerpt sensing score as it discusses the theory behind the imaging technique MRI.
However, the keyword receives a low overall keyword sensing score (Fig. \ref{fig:hm_hardware}) because there are only two excerpts, both in the context of magnetic resonance (Fig. \ref{fig:networkmaps}(a)).
In the quantum computing textbooks, there is no mention of the keyword in Mermin, and only a passing mention in Bernhardt, which vaguely describes the development of NMR machines by companies for magnetic field sensing in the context of quantum engineering.

\textbf{Quantum dot} is an example of a keyword that receives low depth and sensing scores in McIntyre’s \textit{Quantum Mechanics} (Fig. \ref{fig:hm_hardware}).
There is only one excerpt, appearing in the context of quantum wells, which discusses how electron confinement can be used in quantum well diode lasers.
As a result, it receives a low-medium excerpt sensing score and a low keyword sensing score.
In Griffiths, quantum dots are mentioned only once, in a problem in the context of variational principle, and there is no mention of the keyword at all in the modern physics and quantum computing textbooks.

\subsection{Position-first vs Spin-first textbooks}\label{Res:pos_vs_spin}
\begin{figure*}[p] 
    \centering
    \includegraphics[width=\textwidth,height=\textheight,keepaspectratio]{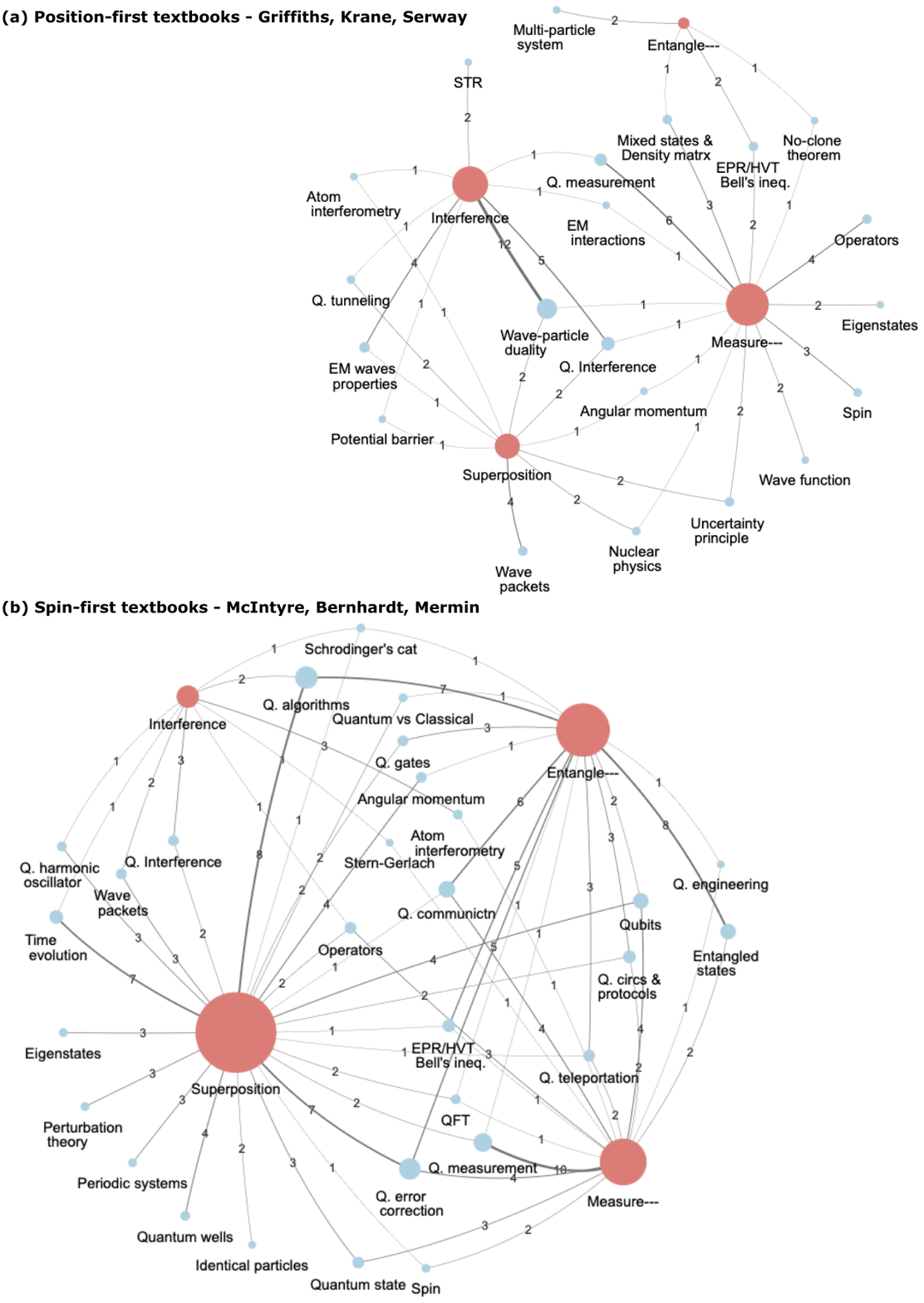}
    \caption{Network maps showing the contexts in which the 4 core keywords appear in (a) the position-first
textbooks (MP books and Griffiths QM), and (b) the spin-first textbooks (QC books and McIntyre QM).  Abbreviations : STR - Special Theory of Relativity, EPR - Einstein–Podolsky–Rosen experiment, HVT - Hidden variable theory, QFT - Quantum Fourier Transform}
    \label{fig:network_s_vs_p}
\end{figure*}

Research question 3 aims to identify differences in how the keywords are addressed and the contexts in which they appear across two types of textbooks: books that use the spin-first approach (McIntyre QM and the QC books) and books that use the position-first approach (Griffiths QM and the MP books). 
To answer this question, we will focus mainly on the 4 core keywords (Superposition, Interference, Entanglement and Measurement) in the interest of clarity and scope.
To visualize the contexts in which these keywords appear across the two types of textbooks, we constructed two separate network diagrams, one corresponding to data from the three position-first textbooks, and the other corresponding to data from the three spin-first (or qubit-first) textbooks (Fig. \ref{fig:network_s_vs_p}).

For better clarity of the visualizations, context nodes of degree 1 (contexts which appear only once for any one keyword) are omitted from the diagrams.
Any edges connected to those nodes are also excluded.
As a result, certain keywords may have appeared in more contexts than what is directly visible in the figure.
We have included the complete version of these network maps in the Supplemental Material \cite{supp}, where we include all the omitted degree-1 nodes, to show that the general trends we observe here hold true.

From the heatmap with the four core concept keywords, we can see that, in general, there are more high-depth discussions on superposition and entanglement in the spin-first books (Fig. \ref{fig:hm_phenomena}), while we see the opposite trend for interference.
Interference appears frequently in the modern physics books, but this is primarily in the context of interferometry experiments in relativity and in wave–particle duality.
Occurrences in the truly quantum sense, for example, interference of quantum states, are fewer in number.
In the spin-first books, almost all occurrences of interference are from McIntyre, including a couple of high depth excerpts, while the quantum computing textbooks hardly address this keyword.
For the keyword measurement, we see medium-high depth excerpts in both the position-first textbooks (mainly in Griffiths) and the spin-first textbooks (in all 3).
We have a detailed discussion of how these 4 keywords are addressed in the books, including their scores and contexts, in section \ref{Res:sec:heatmap_coreconcepts}.

Within the spin-first books, superposition and entanglement appear more often (shown by the node size) and in a wider variety of contexts compared to the position-first books, while interference and measurement appear in a similar number of contexts across the two types of books (Fig. \ref{fig:network_s_vs_p}).
In the position-first books, superposition shows up in the contexts of superposing wavefunctions to form wave packets, wave-particle duality, etc. In the spin-first books, the idea of quantum superposition is addressed more broadly, including how these states behave under time evolution in a variety of systems such as a square potential well, harmonic oscillator, and angular momentum (in McIntyre).
Superposition also appears frequently in the quantum computing books, which address the role of superposition states in quantum algorithms, error-correction protocols, etc.

Regarding entanglement, it is barely discussed in Griffiths and does not appear in the modern physics books at all.
In contrast, the spin-first books offer high-depth discussions of the theory of entanglement and its applications in quantum algorithms and quantum communication, which provide the many contexts seen in Fig.\ref{fig:network_s_vs_p}(b).
In McIntyre, detailed excerpts show up in two chapters: in chapter 4 - Quantum Spookiness, in the contexts of EPR paradox and Schrodinger's cat, and in chapter 16 - Modern Applications of Quantum Mechanics, which contains several QISE-oriented contexts such as quantum algorithms, quantum gates, qubits, and quantum teleportation.
In the quantum computing books, occurrences of entanglement are more widely distributed throughout the books in many contexts including entangled states, EPR paradox, quantum circuits, quantum error correction, quantum teleportation, etc.

Multiple keywords co-occurring within the same context is also substantially more common in the spin-first books (in Fig. \ref{fig:network_s_vs_p}, this is when one context connects to two or more keywords). For example, quantum algorithms, quantum teleportation, EPR/hidden variable theory, quantum measurement, qubits, quantum error correction, and quantum circuits are some of the contexts in which at least three of the four keywords appear simultaneously in the spin-first books.
Several of these contexts with multiple keywords (e.g., qubits) are central to QISE.
In the position-first books, only a few contexts, such as wave–particle duality and quantum interference, include at least three keywords (Fig. \ref{fig:network_s_vs_p}).

Looking at the applications and hardware subsets of keywords, we do not observe any generalizable differences across the spin-first and position-first textbooks (Figs. \ref{fig:hm_applicns}, \ref{fig:hm_hardware}).
However, a clear variation can be seen across the subjects more broadly.
The modern physics books address a wide range of these keywords, not necessarily with a high sensing and depth score, but they touch on different experimental platforms and applications.
For example, imaging, interferometry, resonance, atomic clock, and gravitational wave detector, superconduct--, etc. are some of the application or hardware focused keywords that are addressed in the modern physics books. 
In contrast, the quantum computing books contain very few references to these topics, instead they focus primarily on the theory of QISE.
Most of these keywords do not get mentioned at all in Mermin, and only a few keywords like imaging, superconduct--, and Josephson junction get brief mentions in Bernhardt.
A noticeable difference can be seen between the quantum mechanics books Griffiths and McIntyre.
Griffiths (position-first) contains hardly any mention of these applications/hardware topics, whereas McIntyre (spin-first) addresses many of these keywords (mostly from the Techniques / Applications subset), sometimes with high sensing and depth scores.
McIntyre includes multiple sensing relevant discussions (but low-medium depth) for keywords such as interferometry, atomic clock, gyroscope, and navigation.
Keywords like spectroscopy, NMR, laser cooling, and resonance, are addressed in high depth excerpts that are highly relevant to sensing, resulting in medium-high sensing scores as well.

\section{Results: Course Analysis}\label{Res:course}

Our earlier results (Sec. \ref{TB_overview}) on sensing keywords within textbooks give us an understanding about how we might make conceptual bridges to quantum sensing topics using common quantum-related courses in physics. However, we also wanted to search more broadly across disciplines for examples of courses that might include quantum sensing or at least include quantum and sensing as separate topics. 

\subsection{Distribution of courses across disciplines}

\begin{figure}[h]
\includegraphics[width=0.45\textwidth]{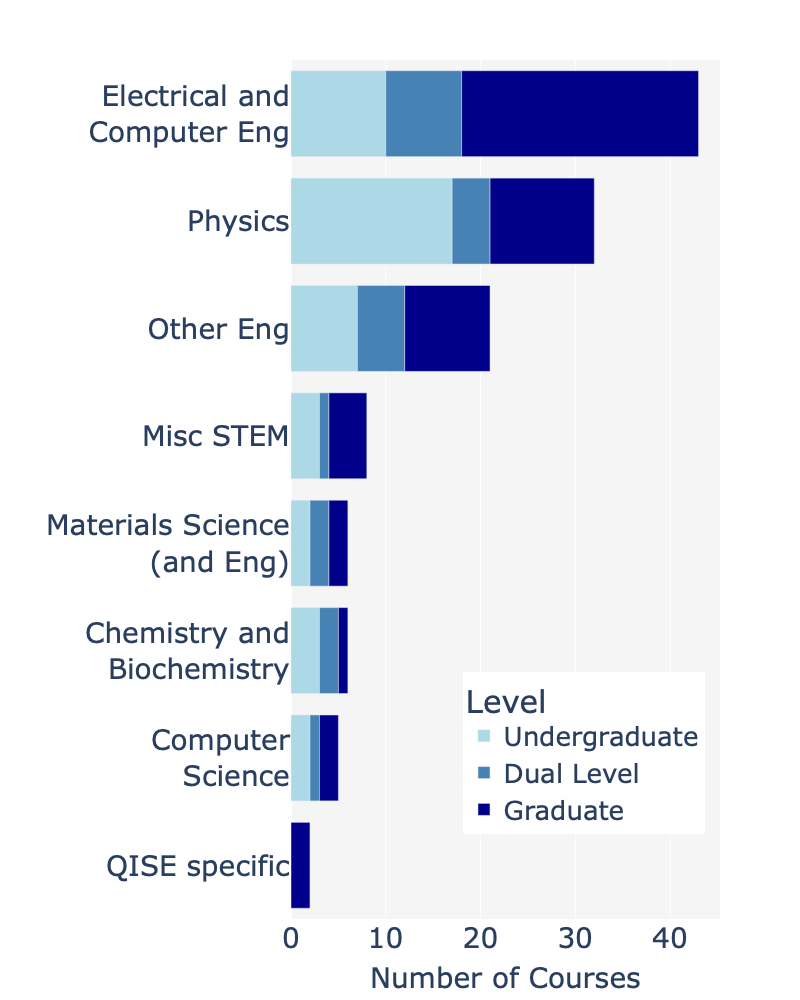}
\caption{\label{fig:disciplines}Distribution of courses (that mention ``quantum” and ``sens-” in the course title and/or description) across disciplines. `Other Eng' refers to other engineering disciplines such as mechanical,
biomedical, engineering science, etc. `Miscellaneous STEM' includes fields like Optical Sciences and
Technology Management}
\end{figure}

Fig \ref{fig:disciplines} shows the distribution by discipline of all courses that mention ``quantum” and ``sens-” in the course title and/or description for undergraduate, dual level, and graduate level courses.
While Electrical and Computer Engineering (ECE) offers the most courses across all levels, Physics has the highest number of undergraduate courses.
The category Other Engineering includes departments such as Mechanical Engineering, Biomedical Engineering, and Engineering Science, while Miscellaneous STEM includes fields like Optical Sciences and Technology Management.
The combined offerings from various engineering departments, including ECE, Materials Science, and Other Engineering, contribute around $55 \%$ of courses that mention ``quantum'' and ``sens''. 

There are two graduate level courses that are cross-listed across multiple departments---one in ECE and Materials Science and the other in Physics and ECE.
In the bar chart, both of these courses have been counted twice, once each in each of their respective departments. So although the total number of courses is 121, the number of courses in the bar chart would add up to 123. 

\subsection{Categorization and analysis of course content}

In this section, we list the 12 categories of courses identified based on a thematic analysis of course content and present descriptions of each category.
For each category, we will list the common course titles, identify the key topics covered, analyze the sensing and quantum content, and identify some sensing-related keywords that are addressed in the course descriptions.
We will also examine how each category addresses other topics directly relevant to QISE, such as quantum computing, information, communication, simulations, cryptography, engineering, etc.
It is important to note that these conclusions are based solely on course descriptions, so they do not fully reflect the course content.

\begin{figure}[h!]
\includegraphics[width=0.45\textwidth]{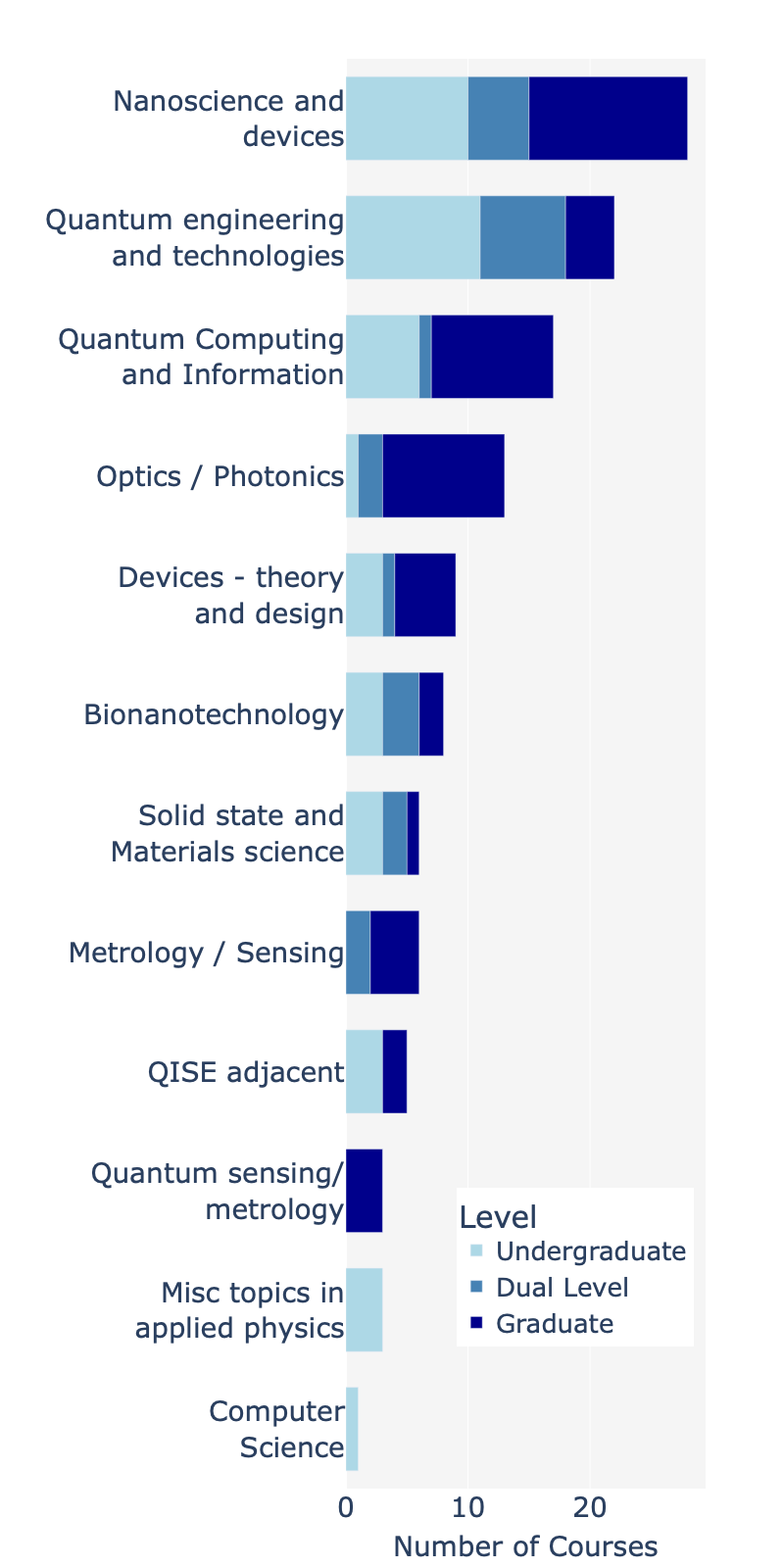}
\caption{\label{fig:categories}Distribution of courses across categories}
\end{figure}

Fig \ref{fig:categories} shows the distribution of all the courses across these different categories, for the three different course levels. The implications of this data are addressed in the discussion/conclusions and reference this section.

\subparagraph{Nanoscience and devices:} The most common course titles within this category include Nanotechnology and Nanoscience and Technology.
A smaller subset of courses have titles such as Nanostructures, Nanophotonics, and Nanomaterials.
The primary focus of these courses is on nanomaterials, including their fabrication, properties, and applications.
Another significant area of emphasis is nanosensors, particularly those involving quantum dots and microelectromechanical systems (MEMS).
In addition, topics related to light-matter interactions and microscopy are covered in select courses.

These courses present sensing-related content to a reasonable extent, offering some but not extensive discussion.
Nanosensors are a key topic in most courses (including a dedicated course on the subject), highlighting the use of nanostructures to develop high-performance sensors.
Sensing-related keywords such as sensor/sensing, quantum dots, photonics, imaging, and detection are among the most frequently found words in the course descriptions.
Various sensor types are addressed, such as plasmonic and photonic crystal sensors, alongside the application of single-electron transistors (SETs) and Micro-Electromechanical Systems (MEMS) in sensor development.
Some courses further explore the role of nanomaterials in imaging techniques.
Although explicit QISE content is minimal, these courses are rich in quantum topics.
This includes discussions on the applications of quantum dots, nanoelectronics, optoelectronics, and nuclear magnetic resonance (NMR), most of which can be considered QISE-adjacent.

10 out of 28 courses in this category are offered exclusively at the undergraduate level, 13 are offered at the graduate level, and the remaining 5 courses are dual level courses (offered to both undergraduate and graduate students).

\subparagraph{Bionanotechnology:} Bionanotechnology and Nanomedicine are common course titles in this category.
The primary foci are nanoscience principles including quantum dots, medical applications of nanomaterials, such as drug delivery, tissue engineering, and DNA sensors.
A few courses address nanosensors, nanobiosensors, and biofuel cells. 

While there is a moderate level of coverage of sensing content in the Bionanotechnology courses, it is not the main focus in any of them.
The course descriptions include keywords such as sensor/sensing, quantum dots, and imaging.
Topics like nanobiosensors, DNA sensors, immunosensors, etc. are mentioned, with an emphasis on using nanomedical tools for diagnostic purposes.
A few courses introduce molecular imaging techniques.
Despite the lack of explicit QISE content, quantum topics are present, primarily through quantum dots, nanomaterials, foundational principles of nanotechnology and its applications in biomedical engineering. 

3 out of 8 courses in this category are undergraduate courses, 2 are offered at the graduate level, and 3 courses are dual level. 

\subparagraph{Quantum Computing and Information:} Courses in this category are typically titled Quantum Information or Quantum Computing, or some variations of these.
The primary focus is on the core quantum information science (QIS) topics, such as quantum computing, quantum networking and communication, or quantum sensing, with significant attention given to quantum states, quantum measurement, and qubits.
Some courses also explicitly mention entanglement and quantum algorithms.  

Sensing-related content in this category is limited, with quantum sensing only mentioned briefly as one of the applications of QIS.
Based on the course descriptions, it appears that none of the courses provides an in-depth exploration of sensing.
The sensing-related keyword that appears frequently in the course descriptions is entanglement, while there are only sparse references to words like metrology and detection.
The only explicit connection to sensing is made in a couple of courses that address topics such as photonic quantum information processing and homodyne detection, but are not central to the course focus.
While the QISE content is obviously substantial, there is minimal QISE-adjacent material, indicating a strong concentration on core QIS concepts.

More than half of the courses in the category are graduate level courses (10 out of 17), 6 courses are undergraduate courses, and only 1 course is offered as dual level. 

\subparagraph{Quantum engineering and technologies:} Most of the courses in this category are titled Quantum Engineering or Quantum Technologies.
The primary focus is on the hardware platforms for quantum devices and the physical implementation of qubits.
Various types of qubits, such as spin-based and superconducting qubits, are explored to some depth.
Entanglement and quantum algorithms are also covered in some courses, typically from the perspective of their role in the development and functioning of quantum devices.
These course descriptions reflect an emphasis on the engineering aspects of quantum systems (including quantum sensors) rather than purely theoretical discussions.

Sensing-related content in these courses is minimal.
Although keywords like sensor/sensing appear multiple times in the descriptions, they seem to show up as applications and are not the key focus of these courses.
Other terms such as entanglement and trapped ions are also mentioned occasionally.
Brief references to sensing-related content include mechanisms such as the operation of atomic clocks, and the role of entanglement in enhancing sensor precision to approach quantum mechanical limits.
Nevertheless, the heavily covered QISE topics like quantum measurement, noise, entanglement, and the physical implementation of quantum hardware are highly relevant to quantum sensing.

Most courses in the category are accessible to undergraduate students - 11 out of 22 are exclusively undergraduate level, 7 courses are dual level, and the remaining 4 are graduate courses.

\subparagraph{QISE adjacent:} This category includes courses covering a range of QISE-adjacent topics that span various disciplines (physics, chemistry, ECE, computer science, etc.), and some of the course titles are Ultracold Quantum Physics, Spin Physics, Quantum Material Spectroscopy, and The Quantum World.
Courses were classified into this category when their primary focus was on quantum-related topics with direct implications to QISE.
Light-matter interactions, quantum theory, and quantum materials are some areas of focus.
Some courses also include topics such as atomic and molecular physics, quantum optics, quantum gases, and spectroscopy. 

Sensing-related content in these courses is limited, with only passing mentions in the descriptions.
Keywords like sensing appear a few times in the course descriptions, especially in the context of optical sensing, homodyne and heterodyne detection, and brief references to quantum technology applications.
Topics such as nanoscale quantum sensing and imaging of quantum materials are also introduced in some courses.
Some foundational aspects of QIS such as quantum communication and quantum technologies get brief mentions.

There are a total of 5 courses identified in this category, 3 of which are undergraduate and 2 are graduate courses.

\subparagraph{Quantum sensing / metrology:} Although this category contains only three courses (all graduate-level), it warrants its own classification, as these courses focus exclusively on quantum sensing and metrology, making them valuable for analyzing course content in this area.
The three courses are titled Quantum Sensing: Machine Learning, Inference and Information; Optical and Quantum Metrology; and Quantum Measurements and Metrology.
As the titles suggest, the main focus areas are sensing mechanisms, quantum measurement, and inference. 

Sensing / sensor, metrology, imaging, spectroscopy, entanglement, and atomic clock are the sensing-related keywords seen in the course descriptions.
The courses delve into the role of information and machine learning in quantum sensing, quantum metrology with a focus on optical and atomic systems, the role of entanglement in measurements, and applications such as timekeeping and imaging.
Concepts such as noise, interferometry, and spectroscopy are also addressed.
All three courses seem to offer comprehensive coverage of quantum sensing and metrology, but not any other QISE topics like quantum computing.

\subparagraph{Devices - Theory and Design:} Courses in this category have titles such as Physics of Electrical Engineering, Mechatronics, Optomechanics, Photovoltaics etc.
The primary focus is on device fabrication, detectors, light-matter interactions, and engineering applications of optical devices, imaging, and noise.
Some courses also discuss MEMS and other sensing mechanisms.
These courses emphasize the engineering and technological aspects of sensing. 

Sensing-related content in these courses is notably high (not always quantum), with frequent mentions of keywords such as sensor, sensing, imaging, and detector.
Topics include optical sensors, transducers in mechatronic systems, noise in electronics, and detector fabrication.
Some courses cover remote-sensing systems, accelerometry, and thermometry, highlighting the broad scope of sensing applications.
Despite the strong emphasis on sensing, there is minimal focus on QISE topics more broadly, as these courses are centered on engineering devices rather than quantum technologies. 
Some of the quantum-related topics mentioned in a few descriptions include quantum engineering, quantum dots, quantum limits/efficiency, and lasers.

Most courses in this category are at the graduate level (5 out of 9), 3 courses are undergraduate level, and 1 course is offered as dual level.

\subparagraph{Solid state and Materials science:} Courses in this category are typically titled Materials Science, Solid State Devices, and Semiconductor Physics.
The primary focus is on semiconductor devices, materials properties, and materials design, with light-matter interactions also playing a significant role.
Some courses delve into nanomaterials, focusing on the intersection of materials science and emerging technologies.
The curriculum emphasizes the study of materials at both the macroscopic and molecular levels, with applications extending to various device technologies.

There is a moderate level of sensing-related content in these courses, often naming applications without additional details.
Sensing-related keywords such as sensor and detector appear regularly, with occasional references to quantum dots and nanoscale concepts.
Related topics include the influence of material properties on device sensitivity, photodetectors, nanoelectronics, and the role of optoelectronics in biological and chemical sensors.
These topics highlight the relevance of material science in the development of sensing technologies.
While QISE content is absent, there is some quantum content, including discussions on optoelectronics, nanotechnology, and quantum materials.

Only 1 of the 6 courses is a graduate level course, the remaining being either undergraduate (3 courses) or dual level (2 courses).

\subparagraph{Metrology / Sensing:} Some of the courses in this category are Physics of Measurement, Modern Imaging, Optical Sensing, and Advanced Biosensors.
Biosensors are a primary focus in many of these courses, with additional emphasis on optical sensing, imaging techniques, and quantum dots.
The courses covers a range of topics that explore the interplay between sensing technologies, methodologies, and applications.

Sensing-related content in these courses is exceptionally high (not always quantum), as evidenced by frequent mentions of keywords such as sensor/sensing, imaging, detection, quantum dots, estimation, and quantum imaging.
Specific topics include biosensors, particularly nanosensors and molecular systems, along with spectroscopy and optical sensing techniques that focus on interferometry, atomic clocks, and remote sensing.
Imaging techniques, such as various scattering methods, point-source imaging, and estimation theory, are also covered in some of the courses.
One course focuses on the theory of fluorescence and its applications in nanostructures and biomedical systems.
While there is no explicit QISE content, mentions of ``quantum'' appear in relation to nanomaterials, quantum measurement, quantum dots, etc.

In this category, there are no courses offered exclusively at the undergraduate level. 4 out of 6 courses are graduate level and the remaining 2 are offered as dual level courses.

\subparagraph{Misc topics in applied physics:} This category includes three courses focusing on some diverse physics topics: Survey of Applied Physics, Experimental Physics, and Modern Physics for Engineers.
These courses do not have a particular area of focus but cover a broad range of topics, including light-matter interactions, interference, information theory, detectors, and signal detection.
The content is geared toward providing a general overview of applied and modern physics concepts, with some courses touching on experimental techniques and their applications.

Sensing-related content in these courses is limited, with only occasional mentions of keywords like sensor and detection.
The Survey of Applied Physics course discusses topics such as quantum computing, materials science, and biophysics, and includes a small emphasis on device and sensor development.
The Experimental Physics course has a bit of sensing-related content like signal-averaging and phase-sensitive detection, while the Modern Physics for Engineers course only covers general modern physics concepts.
While there is little explicit QISE content, other quantum topics such as light-matter interactions, NMR, quantum interference, and wave-particle duality appear more frequently.

All 3 courses in this category are undergraduate level courses. 

\subparagraph{Optics / Photonics:} Courses in this category are primarily titled Photonics, with variations such as Photonics Engineering and Nano \& Micro Photonics.
Some other course titles include Optoelectronics, Information in a Photon, and Coherence Optics.
The primary focus is on applications and methodologies in photonics, optical sensing, optical devices and circuits, and light-matter interactions.
A few courses also touch on topics like semiconductor devices, interference, and quantum dots, broadening the scope of photonics to related fields.

Sensing-related content in these courses is above average, with frequent mentions of keywords such as sensing, photonics, detector/detection, imaging, quantum dots, and interference.
The main sensing-related focus is on optical detection, including interferometric manipulations, optoelectronic detection systems, imaging, and quantum-enhanced detection.
Other topics include light detection in semiconductors, optical communication and sensing, and the design and development of optical circuits and devices.
One particular course, Photonics Devices and Sensors, has a strong emphasis on sensing and explores a wide range of technologies, from lasers, cameras, computer displays, to solar cells and molecular sensing.
While explicit QISE content is minimal, the courses heavily feature some quantum-adjacent topics like photonics and optoelectronics, reflecting their close relationship with quantum technologies.

Most courses in this category (10 out of 13) are graduate level courses. 2 courses are offered as dual level and only 1 out of 13 is an undergraduate course.

\subparagraph{Computer science:} This category has only one course taught at the undergraduate level, called Parallel Computer Architecture, the focus area of which is multiprocessor architectures. Other topics include interconnection networks, CPU design, and design of large-scale storage systems. 
The course description briefly mentions quantum computing and wireless (non-quantum) sensor networks, and this seems to be the only sensing-related content covered.

\section{Discussion and Conclusion}

As mentioned in the background section, in a prior study examining the landscape of quantum-related courses, a large fraction of the approximately 4700 physics courses that mention ``quantum” were found to be quantum mechanics ($\sim$1600) and modern physics ($\sim$600) courses \cite{pina_landscape_2025}.
Also, the majority of dedicated QISE courses (418 out of 514) identified in the dataset were focused on quantum computing.
This suggests that modern physics, quantum mechanics, and quantum computing courses have a broad curricular reach, especially in physics.
However, our course analysis reveals that coverage of quantum sensing is very rare in both QISE courses and the widely taught quantum mechanics and modern physics courses.
However, because these courses already provide some foundational quantum concepts, they have a strong potential for the integration of quantum sensing (QS) concepts.
Moreover, our textbook analysis shows that each of these three subjects includes certain topics that can serve as entry points for incorporating sensing-related ideas, as discussed later in this section.
Overall, our study focuses on how sensing-related concepts are presently addressed in these three subjects, which will enable educators and curriculum developers to identify opportunities for integrating quantum sensing within the current curriculum. We believe that integration with the existing curriculum is more feasible, especially at the undergraduate level, than creating entirely new courses on quantum sensing, which is just one of multiple important topics within QISE and quantum technology. 

\subsection{Textbook analysis}

\subparagraph{RQ 1 and 2:} Research questions 1 and 2 examine how the three subjects differ in their treatment of sensing-related ideas and the contexts in which these ideas appear.
The two modern physics (MP) textbooks show a similar coverage of topics.
They tend to weave in a range of hardware and application-focused topics.
However, these discussions are often surface-level and do not delve into the underlying physical ideas.
Some exceptions to this include the discussions on NMR in Serway and superconductivity in both Krane and Serway, which provide more comprehensive discussions.
Compared to the other subjects, MP textbooks also cover a broader range of topics, including non-quantum contexts, which may offer opportunities to integrate multiple concepts that build towards some central idea.
At the same time, MP books lack in-depth treatment of the core concepts (superposition, interference, entanglement, and measurement), especially entanglement, which is absent altogether.
This represents a clear gap that educators and curriculum developers could address, as introducing such core concepts more rigorously could help students build a stronger foundation for advanced quantum and QISE topics.

In contrast, both the quantum computing (QC) books provide a more integrated and detailed treatment of the concepts of superposition, entanglement, and measurement.
While these appear in other subjects as more isolated topics, the QC books address the interplay of these topics and their application in contexts like quantum algorithms, quantum error correction, qubits, etc.
The QC books also include various QISE-relevant contexts, and this, along with the detailed treatment  of the core concepts, can be leveraged to introduce more advanced QISE topics like quantum sensing.
However, the QC textbooks examined in our study lack discussions on hardware and applications, and they primarily focus on theoretical concepts of quantum information science.
QC books present an abstract theory that applies to any hardware, which may explain why these hardware-focused topics are absent.
In addition, it can be challenging to incorporate topics like sensing applications because quantum sensing lacks a similar hardware-agnostic conceptual framework, and hence they may be beyond the scope of the textbooks.

We see a notable difference between the two quantum mechanics (QM) textbooks, unlike MP and QC.
While Griffiths does not include many detailed discussions of core QISE ideas like quantum superposition and entanglement, McIntyre provides high depth explanations of these concepts in more than one context.
McIntyre also covers hardware related topics like NMR (more detailed) and quantum dots (briefly), and experimental techniques such as laser cooling and interferometry, particularly alongside mentions of sensing topics like atom interferometers and atomic clocks.
This contrast in the content of the books suggests that, if an instructor's goal is to link standard QM topics with QISE topics, including QS, then it might be beneficial to use McIntyre instead of Griffiths.

\subparagraph{RQ 3:} Research question 3 addresses how textbooks that use different approaches (spin-first vs position-first) differ in their treatment of the sensing-related keywords.
A major difference between the two types of books lies in how they address superposition and entanglement.
The spin-first books (QC and McIntyre's QM) cover these topics a lot more compared to the position-first books (MP and Griffiths's QM), both in terms of the excerpt depth and the number of excerpts.
Although the position-first books include discussions of superposition, these do not necessarily deal with the concept in a purely quantum sense, but it may come up in other contexts, such as wave packet formation.
While entanglement is a central idea in the spin-first books (especially in QC, but also addressed extensively in McIntyre), it hardly shows up in the position-first books (a few excerpts in Griffiths only).
The few excerpts that are present appear mostly in the context of multi-particle systems, where the concept is introduced, and EPR paradox, which takes a philosophical and theoretical approach.
Additionally, the position-first books did not include QISE-relevant applications of entanglement, such as quantum teleportation or quantum algorithms.
Generally, the spin-first books include a much broader coverage of QISE topics, and hence these might provide an easier bridge to quantum sensing. 

Another important aspect worth reflecting on is the contexts in which the core concepts (superposition, interference, entanglement, interference) appear together.
Since these concepts are closely related, their joint treatment within the same discussion can be indicative of a more coherent and better integrated QISE curriculum.
While the position-first books have very few contexts with co-occurring keywords, the spin-first books address some of these concepts simultaneously in areas such as quantum algorithms, quantum teleportation, qubits, quantum error correction, EPR/HVT, and more.
These differences suggest that a particular choice of QM textbook (and consequently spin-first vs position-first approach) could have implications in regards to preparation for QISE topics.
The distinction is not only about which formalism students encounter first, but also how core concepts are framed, what topics are covered, the depth of coverage, and how QISE topics might be integrated. Some of this is difference is historical, as the first edition of Griffith's QM textbook is from 1995, which was still early days of quantum computing research. Quantum mechanics textbooks from that era had different priorities for applying the formalism and concepts of QM. 

\subparagraph{Insights for teaching quantum sensing:} This study helps identify areas within the existing curriculum that can serve as bridges to quantum sensing content.
One type of these `bridging’ topics includes cases where we can draw direct analogies between topics covered in these textbooks and quantum sensing topics.
For example, the double-slit experiment (which appears in both MP and QM textbooks) and the Michelson-Morley experiment (which is extensively addressed in MP) can be used to introduce the concept of interferometry.
We can build on these discussions to introduce ideas on general interferometric techniques and how interferometric phase measurement yields insight about a quantity to be sensed.
Instructors can further elaborate on these concepts to introduce advanced ideas like atom interferometry (briefly addressed in McIntyre) or Ramsey interferometry used in atomic clocks, provided that the students have sufficient familiarity with the core concepts like superposition.
Another bridge would be the discussions on nuclear magnetic resonance (NMR) in Serway and McIntyre, since direct analogies can be drawn between pulse sequence control in NMR and Ramsey interferometry.

Another way to introduce sensing topics is by building on specific topics that are covered in the curriculum.
For instance, both MP and QM textbooks cover the Zeeman effect, which is the foundational idea used in NV center magnetometry.
Instructors could potentially introduce the NV center sensing protocol as an application of concepts such as Zeeman effect, resonance, light-matter interaction, etc.
Another example appears in the MP textbooks, which address the uncertainty principle in significant detail and also introduce the idea of gravitational wave detection, but as isolated topics.
This provides an opportunity to integrate these ideas and bring in the concepts of squeezed light and its application in enhancing the detection sensitivity of LIGO (Laser Interferometer Gravitational Wave Observatory).
More broadly, entanglement is substantially covered in the QC books and McIntyre’s QM.
This can serve as a strong foundation for introducing concepts such as quantum advantage or entanglement-enhanced sensitivity of quantum sensors. 

Finally, there are a few direct references to sensing platforms in these textbooks, but accompanied by brief superficial discussions only.
One example is the brief mentions of atomic clocks in the MP books (as a tool in relativity experiments) and in McIntyre (in the context of hyperfine transitions and laser cooling).
Other examples include quantum dots in McIntyre and superconducting quantum interference devices (SQUIDs) in Krane, both brief discussions.
These excerpts could be used to introduce students to the detailed sensing protocols and underlying physics of these platforms. 

As discussed earlier in this section, it will also be helpful for QM instructors to be mindful of the differences between the spin-first and position-first textbooks.
The choice of the textbook could influence not only the mathematical formalism used, but also the extent to which QS related topics can be incorporated into the course.

\subparagraph{Limitations:} A major limitation of the study is in how we are calculating the sensing and depth scores for each keyword.
For example, as mentioned in the score calculation section (section \ref{subsec:Method_rubric_heatmap}) , looking only at the total sum for sensing scores could be misleading while interpreting the results.
For instance, a keyword with many excerpts, each with low sensing content, could end up with a higher total score than a keyword with just one excerpt that provides a rich discussion.
Or alternatively, a keyword with a single high sensing score excerpt might still appear to have a low overall keyword sensing score.
We did consider several methods for score calculation, including average, best excerpt, weighted sum, etc. While each approach has its own strengths and weaknesses, we selected the total sum for sensing and the best excerpt for depth, as we felt these metrics provided the most interpretable and meaningful results for our analysis.

Our list of keywords seem to cover most of the relevant material, with some important sections even being tagged multiple times.
However, the list is not exhaustive.
Future work could involve expanding the set of keywords and including more textbooks from other disciplines (engineering, chemistry, etc.) for analysis.

\subsection{Course analysis}

\subparagraph{RQ 4 and 5:} Most courses that mention ‘quantum’ and ‘sens-’ are offered within engineering departments, with ECE and other engineering fields together contributing a substantial number of courses, particularly at the graduate level.
Physics has the largest share of undergraduate courses.
The majority of these courses fall under the category of nanoscience or nanotechnology, most of which address the use of nanostructures such as quantum dots and MEMS for sensing applications.
Notably, the categorization also shows that there are very few dedicated courses in quantum sensing / metrology (N=3), none of which are at the undergraduate level.

\subparagraph{Implications:} The fact that only 121 courses mentioned ‘sens-’ out of 8456 courses in the dataset indicates that quantum sensing is a low priority in current courses.
Furthermore, the lack of dedicated quantum sensing courses at the undergraduate level highlights the need for curriculum development efforts.
Since physics accounts for the largest number of undergraduate quantum-related courses with a sensing component, it is especially relevant to focus improvement efforts on the undergraduate physics curriculum (especially in the canonical quantum courses like QM and MP).
Our textbook analysis offers one approach to connecting existing quantum-related undergraduate curricula to applications in quantum sensing.

\subparagraph{Limitations:} A major limitation of our course analysis is that we look only at course titles and descriptions.
Since course descriptions may not always provide a fully accurate, comprehensive, or up-to-date reflection of the course content, the results of this analysis should be interpreted with caution.
To address this, future work could include more extensive data sources such as syllabi or instructor interviews.

Although our dataset does not include every course offered in the U.S., it draws data from the larger quantum education landscape study, which encompasses data from over 8,000 courses across all departments at more than 1,400 institutions (including all R1 and R2 institutions, all minority serving institutions, all institutions with accredited engineering and computer science degrees, and several of the highest STEM degree granting institutions within each state).
Given the large scope of this dataset, we can be fairly certain that our sample represents a substantial majority of all quantum-related courses in the U.S.

The 12 categories were identified solely based on course descriptions and similar prior work done as part of the quantum education landscape study.
Additionally, the data used for this analysis was collected in Fall 2024, and course offerings may have changed since then through revisions to existing courses or the introduction of new ones.

\section{Acknowledgments}

We would like to thank Dr.\ Andi Pi\~{n}a for their valuable guidance and support throughout this project.
We also thank the members of the Center for Advancing Scholarship to Transform Learning (CASTLE) at Rochester Institute of Technology for their helpful feedback and productive discussions.
This work is funded by National Science Foundation under award 2315691.

\bibliography{Bibliography}

\end{document}


\preprint{APS/123-QED}

\title{Supplemental material - \\ Analysis of quantum-related courses and textbooks for potential integration of quantum sensing}

\author{Namitha Pradeep}
 \email{np6895@rit.edu}
\author{Ben Zwickl}%
\altaffiliation[Also at ]{Center for Computing in Science Education, University of Oslo, Olso Norway}
 
\affiliation{%
 Rochester Institute of Technology, School of Physics and Astronomy, Rochester, NY, USA
}%



\maketitle
\section{Introduction}

In this Supplemental Material document, we present three additional results. In section \ref{sec:keywords_general}, we discuss a few additional keywords that represent general terms used in sensing and metrology.
We include the heatmap and discuss their sensing and depth scores, as well as the contexts in which these keywords appear across the six textbooks.
Section \ref{sec:network_subjects} shows the complete versions of the network maps for the three subjects - modern physics (MP), quantum mechanics (QM), and quantum computing (QC).
These maps include the weight-1 edges that were omitted from the maps shown in the main paper.
Similarly, in section \ref{sec:network_spin_vs_posn}, we present the complete versions of the spin-first vs position-first network maps, where we include all the nodes of degree 1.

\section{Keywords in the General category}\label{sec:keywords_general}

In this section, we will address some additional keywords that belong in the `General' category, which includes general terms used for sensing or detection purposes (Table \ref{tab:keywords_general}). We will discuss the sensing and depth scores of these keywords (visualized in the heatmap - Fig \ref{fig:hm_general}), and also the contexts in which they appear in the textbooks (based on the network maps in the main paper).

\begin{table}[h]
    \centering
    \begin{tabular}{|p{3cm}|c|}
        \hline
        Category & Keywords \\ 
        \hline
        \multirow{10}{3cm}{General} & Quantum technology \\
                                      & Quantum imaging \\
                                      & Quantum sensor \\ 
                                      & Sensor \\ 
                                      & Sensing \\ 
                                      & Metrology \\
                                      & Detector / Detection \\
                                      & Estimation / Estimator \\
                                      & Probe \\
        \hline
        
    \end{tabular}
    \caption{List of keywords in the General category}
    \label{tab:keywords_general}
\end{table}

\subsection{Heatmap : General}\label{subsec:heatmap_general}

\begin{figure*}[hbtp]
\includegraphics[scale=0.95]{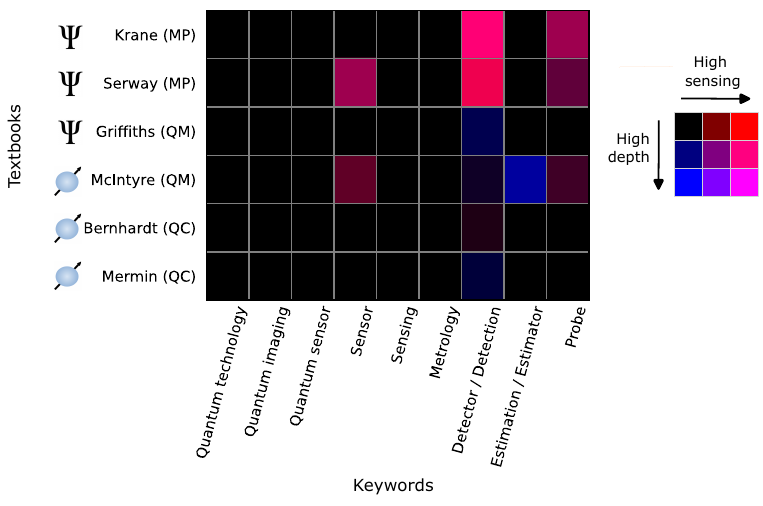}
\caption{\label{fig:hm_general}Heatmap showing the sensing and depth scores for the keyword category - `General terms'. The icons next to the textbook names indicate whether they use a position-first (\includegraphics[height=1.8ex]{psi_figure.png}) or spin-first (\includegraphics[height=1.8ex]{spin_figure.png}) approach. The color mixing scheme used in the heatmap is described in the main paper. Although the legend shows a discrete set of colors, the heatmap itself uses a continuous color scale.}
\end{figure*}

We start with the keyword detector/detection, which has some high sensing and medium depth score examples.
We then look at medium sensing and low depth examples for keywords probe and sensor (see heatmap - Fig. \ref{fig:hm_general}).
Interestingly, none of the other keywords in this category are addressed in any of the textbooks (estimation/estimator being an exception - it gets a very brief mention in McIntyre). 

\textbf{Detector/detection - }An example for high sensing score and a medium depth score is the keyword detector/detection in the modern physics textbooks (Fig. \ref{fig:hm_general}).
There are many (10-20) excerpts tagged under this keyword in both Krane and Serway.
In Krane, one excerpt appears in the subsection `Detecting Particles' within the section on particle interactions and decays, in the context of particle physics (Fig. 2(a) in the main paper).
It includes a couple of equations, a few diagrams, and a rich conceptual discussion, which results in a medium excerpt (and keyword) depth score.
The excerpt discusses different types of detectors, explaining their physical implementation and sensing protocols, and hence receives a high excerpt sensing score.
Other excerpts also cover topics such as gravitational wave detection (context - gravitational waves), photodiodes for sensing (context - semiconductors), and imaging using radioisotopes (context - nuclear physics) (Fig. 2(a) in the main paper), contributing to the overall high keyword sensing score.
In Serway, the section on radiation detectors in the context of nuclear physics (Fig. 2(a) in the main paper) explains the theory and applications of various radiation detectors.
This excerpt receives a low-medium excerpt (and keyword) depth score due to its conceptual discussion and use of a few diagrams.
Since it explains the working principles and implementation of different detection schemes, it gets a high excerpt sensing score.
As in Krane, other excerpts address topics such as gravitational wave detection and particle detectors, contributing to the high keyword sensing score (Fig. \ref{fig:hm_general}).
Additional contexts in which this keyword appears in modern physics include special theory of relativity (STR), atoms (atomic models), properties of electromagnetic waves (energy, power, Compton effect, etc.), wave-particle duality (diffraction experiments), and the uncertainty principle (Fig. 2(a) in the main paper).

In quantum mechanics, there are a moderate number of excerpts associated with the keyword detector/detection, but they receive low sensing and depth scores overall (Fig. \ref{fig:hm_general}).
These excerpts appear in contexts such as the EPR experiment, Stern-Gerlach experiment, scattering (including theory and problems on cross-section experiments), and time evolution (neutrino oscillations) (Fig. 2(b) in the main paper).
In the quantum computing textbooks, the keyword appears in only a few excerpts and receives low scores (Fig. \ref{fig:hm_general}).
It is briefly mentioned twice in the context of quantum measurement (measurement along different bases) (Fig. 2(c) in the main paper), and once in a discussion of quantum error correction, which is not shown in the network diagram.

\textbf{Probe} is an example of a keyword that received a medium sensing score and a low depth score in modern physics textbooks (Fig. \ref{fig:hm_general}).
In Krane, an excerpt on the scanning tunneling microscope, within the context of potential barriers, receives a low excerpt depth score due to the lack of mathematical detail, but a medium excerpt sensing score as it explains the imaging protocol.
The keyword also appears in discussions of photodiodes as detectors and in the measurement of cosmic microwave background (CMB) radiation, contributing to an overall medium-high keyword sensing score.
Serway includes very similar excerpts for the keyword probe and therefore receives similar scores.
The keyword appears repeatedly in the context of cosmology (CMB radiation, general relativity, etc.) across both books (Fig. 2(a) in the main paper), and also appears once in the contexts of semiconductors and quantum tunneling.
In the quantum mechanics textbooks, it does not appear in Griffiths, but is mentioned briefly twice in McIntyre---once in the context of Schrödinger’s cat and once in atom interferometry, earning a low sensing score (Fig. \ref{fig:hm_general}).
The keyword does not appear in the quantum computing textbooks.

\textbf{Sensor - }Similarly, the keyword sensor also receives a medium sensing score and a low depth score in Serway’s \textit{Modern Physics} (Fig. \ref{fig:hm_general}).
There are only two excerpts: one discussing charge-coupled devices in the context of quantum wells, and the other in the context of atom interferometry for gravitational sensing.
Both excerpts receive low depth scores due to the lack of any mathematical discussion but receive high excerpt sensing scores.
Since there are only two excerpts, this results in a medium keyword sensing score.
In contrast, the keyword does not appear at all in Krane.
In quantum mechanics, it does not appear in Griffiths, but is mentioned very briefly twice in McIntyre in the chapter on Periodic Systems: once in the discussion on application of semiconductors (light-emitting diodes and light sensors), and once in the context of biosensing applications of graphene and carbon nanotubes.
It receives a low-medium keyword sensing score and a low keyword depth score (Fig. \ref{fig:hm_general}).
The keyword sensor does not appear in the quantum computing textbooks.

\section{Network maps for MP, QM and QC textbooks}\label{sec:network_subjects}

\begin{figure*}[p] 
    \centering
    \includegraphics[width=\textwidth,height=\textheight,keepaspectratio]{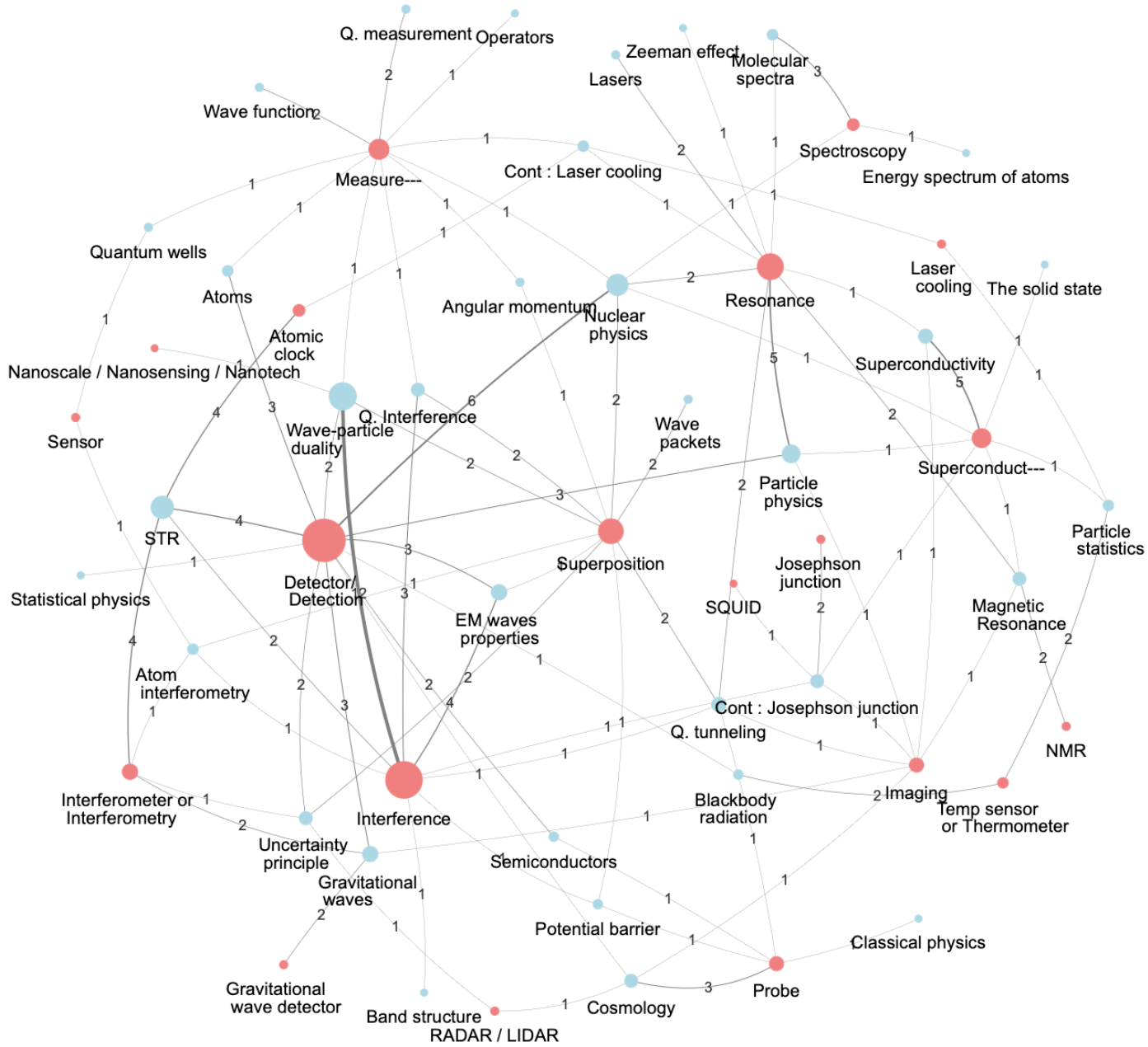}
    \caption{Network map showing the contexts in which the keywords appear in the MP textbooks - Krane and Serway. The red nodes are the keywords and the blue nodes are the contexts. Abbreviations : EPR - Einstein–Podolsky–Rosen experiment, HVT - Hidden variable theory, NMR - Nuclear Magnetic Resonance}
    \label{fig:network_MP}
\end{figure*}

\begin{figure*}[p] 
    \centering
    \includegraphics[width=\textwidth,height=\textheight,keepaspectratio]{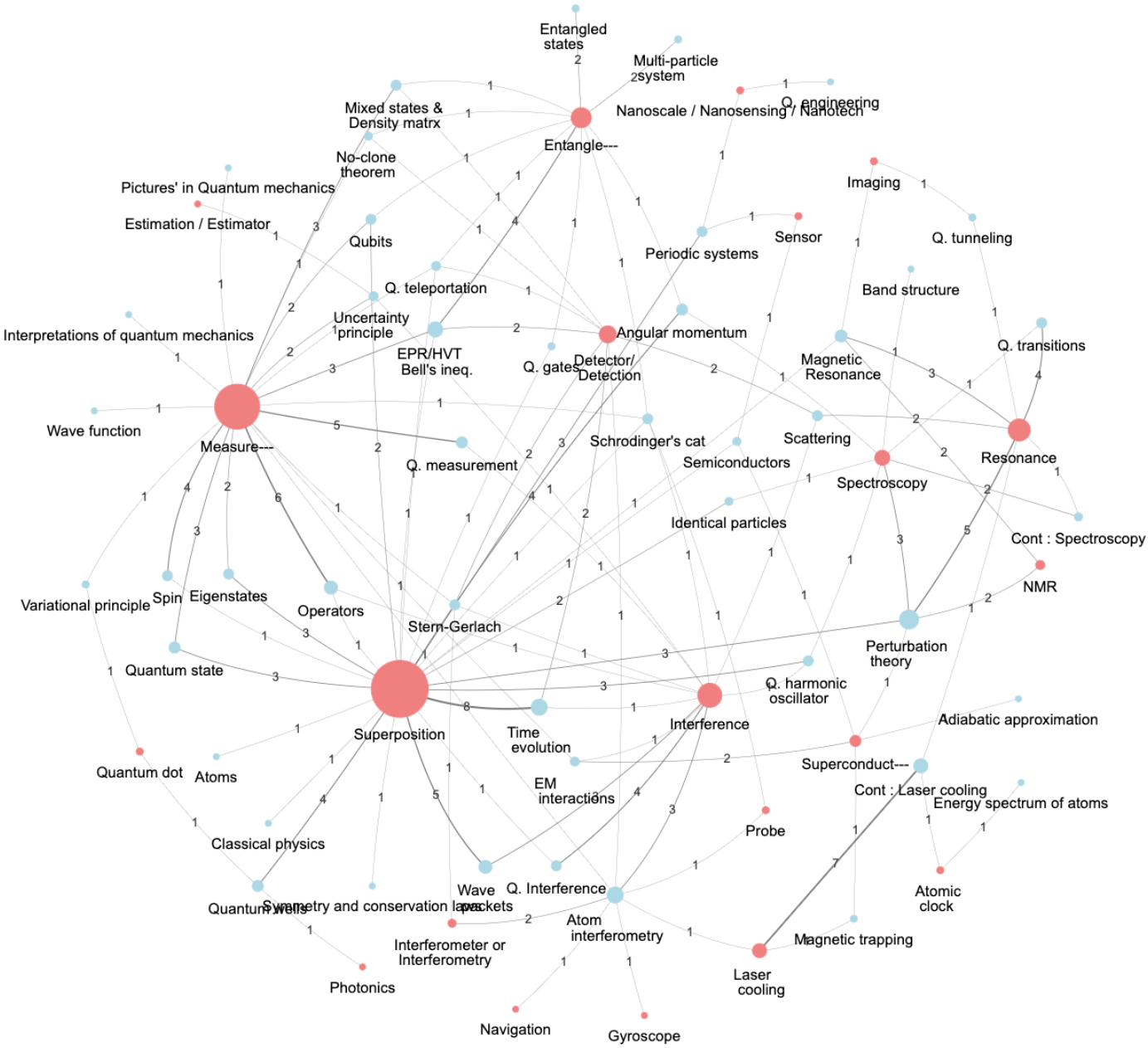}
    \caption{Network map showing the contexts in which the keywords appear in the QM textbooks - Griffiths and McIntyre. The red nodes are the keywords and the blue nodes are the contexts. Abbreviations : STR - Special Theory of Relativity, NMR - Nuclear Magnetic Resonance}
    \label{fig:network_QM}
\end{figure*}

\begin{figure*}[p] 
    \centering
    \includegraphics[width=\textwidth,height=\textheight,keepaspectratio]{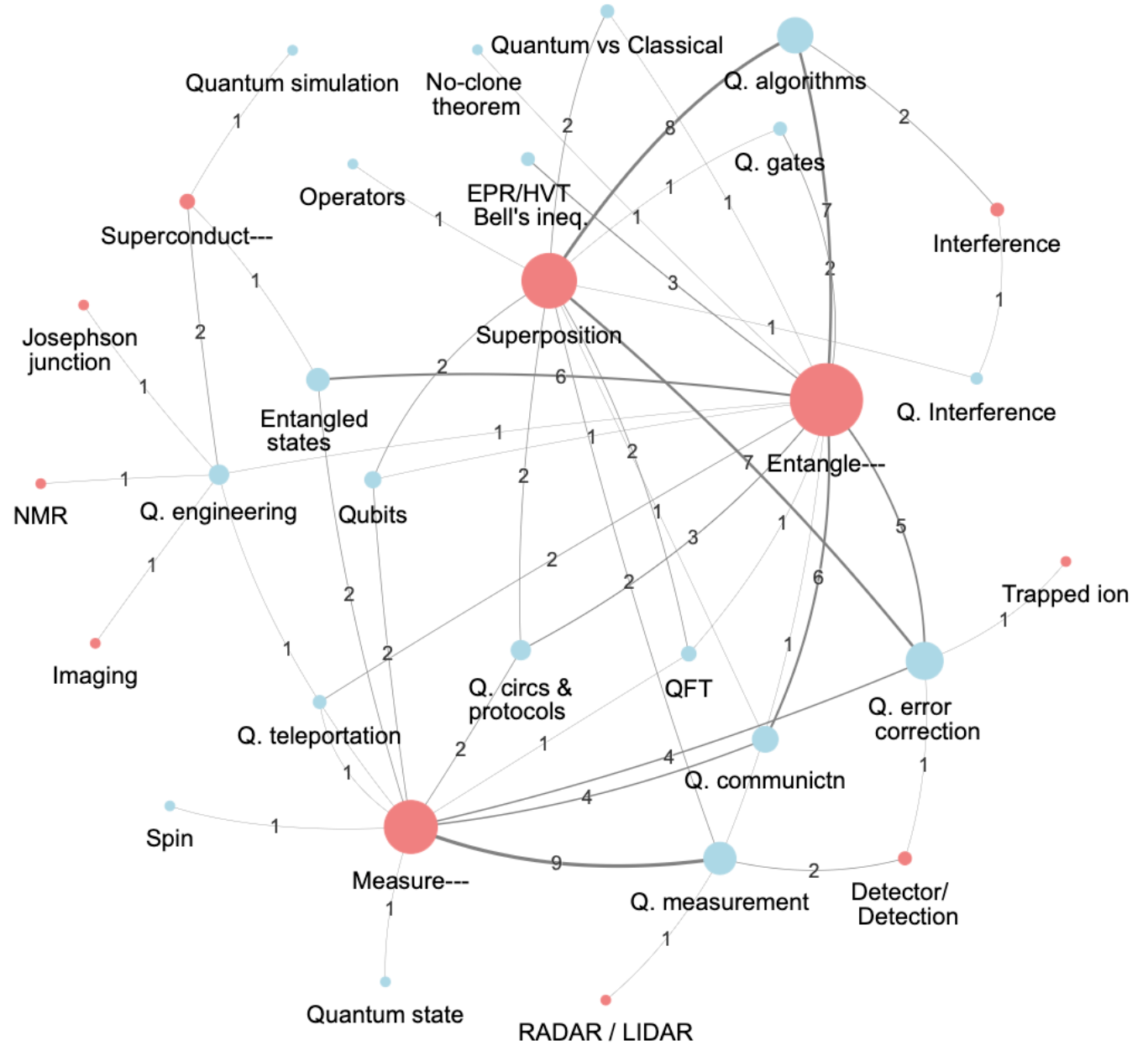}
    \caption{Network map showing the contexts in which the keywords appear in the QC textbooks - Bernhardt and Mermin. The red nodes are the keywords and the blue nodes are the contexts. Abbreviations : EPR - Einstein–Podolsky–Rosen experiment, HVT - Hidden variable theory, NMR - Nuclear Magnetic Resonance, QFT - Quantum Fourier Transform}
    \label{fig:network_QC}
\end{figure*}

In this section, we present the network maps that visualize the contexts in which the keywords appear in the textbooks.
These maps are identical to those shown in the main paper (Fig. 2 in the main paper), but include all weight-1 edges for completeness.
Fig. \ref{fig:network_MP} shows the network map for the modern physics textbooks Krane and Serway, fig. \ref{fig:network_QM} for the quantum mechanics books Griffiths and McIntyre, and fig. \ref{fig:network_QC} for the quantum computing books Bernhardt and Mermin.

\section{Network maps - Spin-first vs Position-first textbooks}\label{sec:network_spin_vs_posn}

\begin{figure*}[p] 
    \centering
    \includegraphics[width=\textwidth,height=\textheight,keepaspectratio]{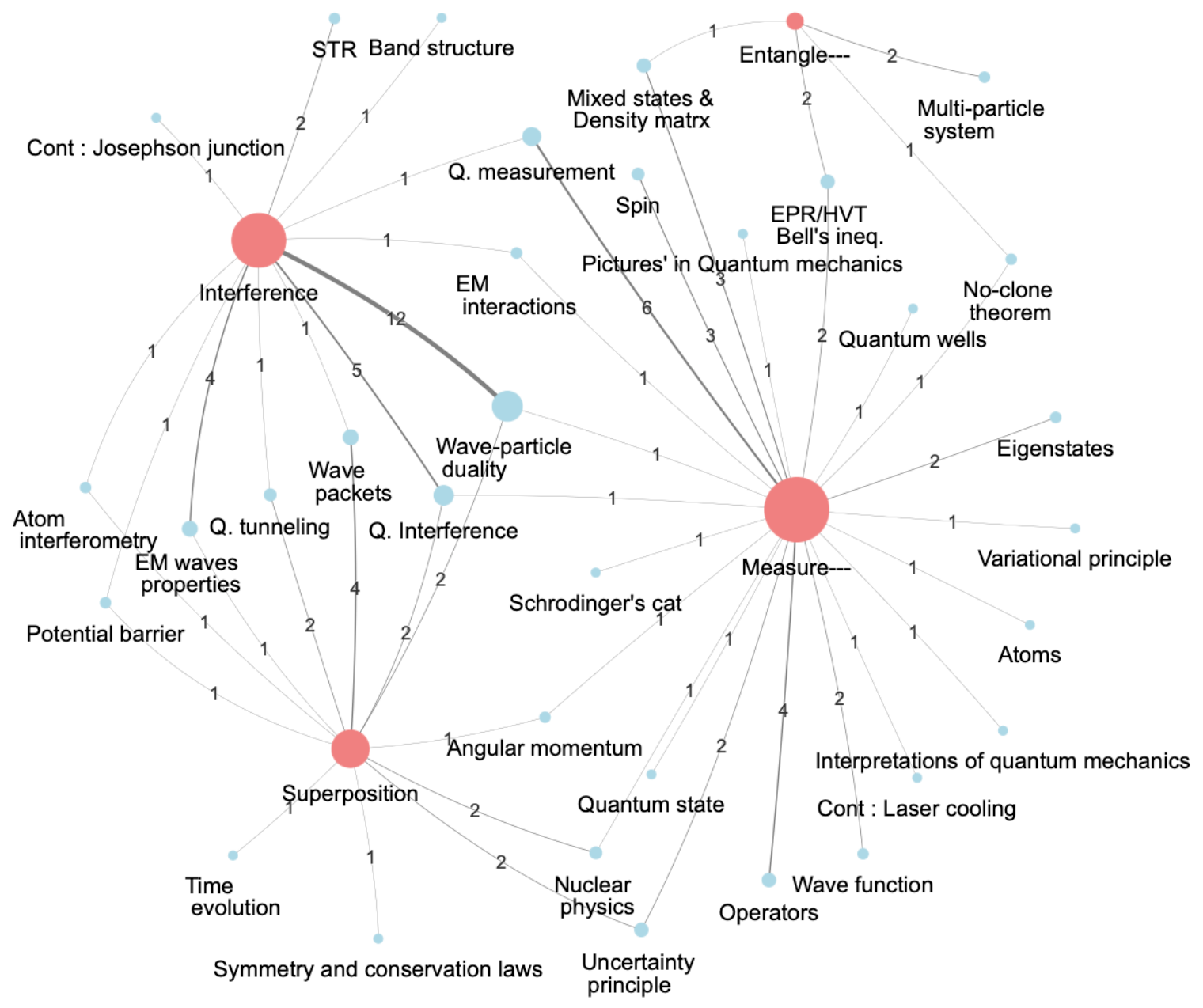}
    \caption{Network map showing the contexts in which the keywords appear in the position-first textbooks - Griffiths (QM), Krane (MP), and Serway (MP). The red nodes are the keywords and the blue nodes are the contexts. Abbreviations : STR - Special Theory of Relativity, EPR - Einstein–Podolsky–Rosen experiment, HVT - Hidden variable theory, EM - electromagnetic}
    \label{fig:network_position}
\end{figure*}

\begin{figure*}[p] 
    \centering
    \includegraphics[width=\textwidth,height=\textheight,keepaspectratio]{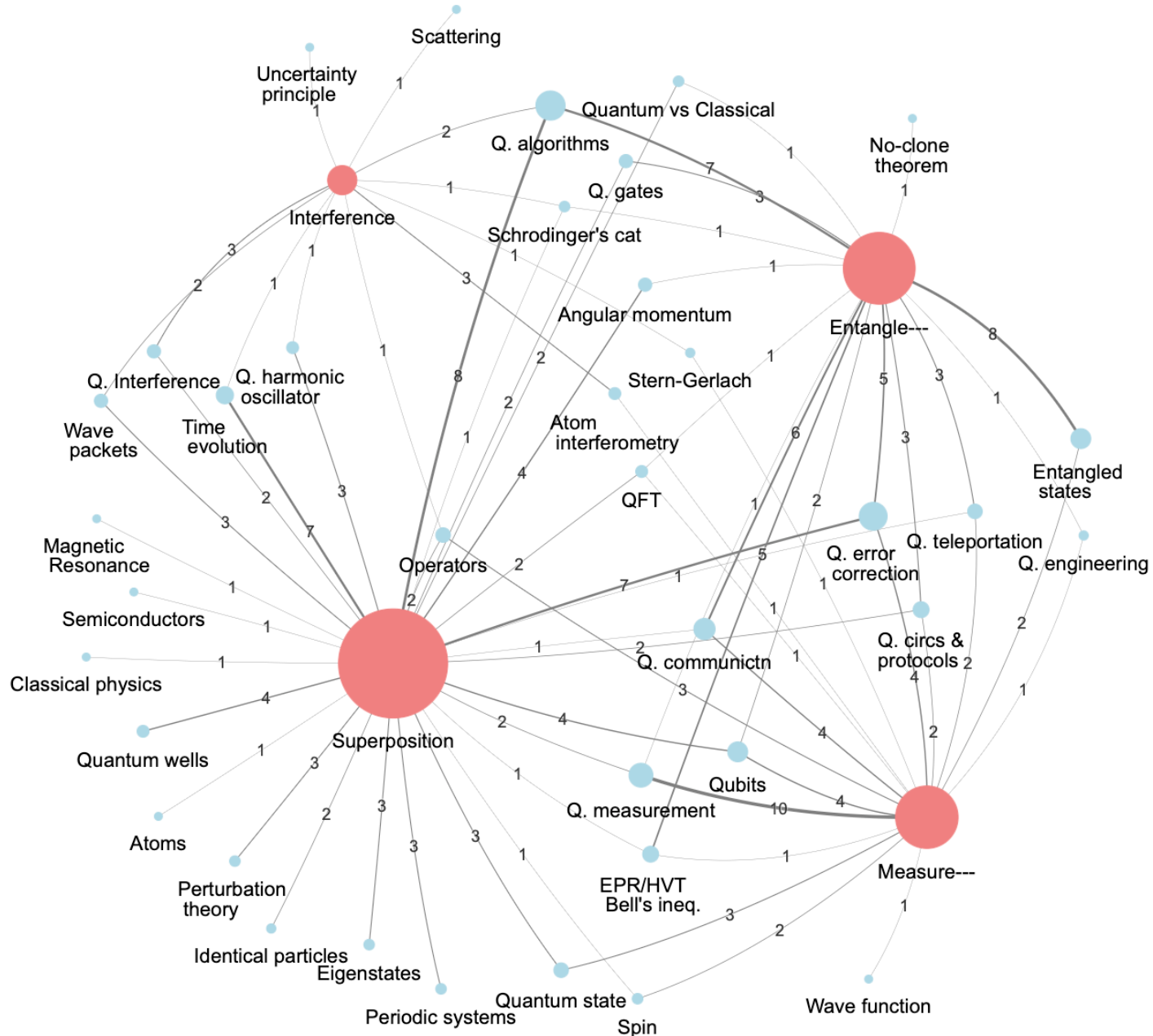}
    \caption{Network map showing the contexts in which the keywords appear in the spin-first textbooks - McIntyre (QM), Bernhardt (QC), and Mermin (QC). The red nodes are the keywords and the blue nodes are the contexts. Abbreviations : EPR - Einstein–Podolsky–Rosen experiment, HVT - Hidden variable theory, QFT - Quantum Fourier Transform}
    \label{fig:network_spin}
\end{figure*}

In this section, we present the network maps that visualize the contexts in which keywords appear in the position-first (Fig. \ref{fig:network_position}) and spin-first textbooks (Fig. \ref{fig:network_spin}).
These maps are identical to those shown in the main paper (Fig. 6 in the main paper), but include all the nodes with degree 1 for the sake of completeness.

